\documentclass[12pt]{article}

\usepackage{graphicx} 
\usepackage{authblk} 
\usepackage{tabularx} 
    \newcolumntype{L}{>{\raggedright\arraybackslash}X}
    \newcolumntype{C}{>{\centering\arraybackslash}X}
    \newcolumntype{R}{>{\raggedleft\arraybackslash}X}
\usepackage{colortbl} 
\usepackage{float} 
\usepackage{placeins} 
\usepackage{amsmath} 
\usepackage{amsfonts} 
\usepackage{bm} 
\usepackage{braket} 
\usepackage{url} 
\usepackage{hyperref} 
\usepackage{pdflscape} 
\usepackage{dirtytalk} 
\usepackage{braket} 
\usepackage[utf8]{inputenc}
\usepackage{siunitx} 
\usepackage{algorithm}%
\usepackage{algorithmicx}%
\usepackage{algpseudocode}%
\usepackage{mathrsfs}   
\usepackage{float}
\usepackage{enumitem} 
\usepackage[T1]{fontenc}
\usepackage{pdflscape} 
\usepackage{subcaption} 
\usepackage{threeparttable} 

\usepackage{cite} 

\begin{document}

    \title{Solving larger Travelling Salesman Problem networks with a penalty-free Variational Quantum Algorithm}     

\author[1, 2]{Daniel Goldsmith \thanks{Corresponding Author: D.Goldsmith@kingston.ac.uk}}
\author[1]{Xing Liang \thanks{Principal Corresponding Author: X.Liang@kingston.ac.uk}} 
\author[1]{Dimitrios Makris}
\author[3]{Hongwei Wu}

\affil[1]{Faculty of Engineering, Computing and the Environment, University of Kingston, London, KT1 2EE}
\affil[2]{Digital Catapult, 101 Euston Road, London, NW1 2RA}
\affil[3]{Centre for Engineering Research, University of Hertfordshire, Hatfield, AL10 9AB}

\date{September 2026}

\maketitle

\begin{abstract}
The Travelling Salesman Problem (TSP) is a well-known NP-hard combinatorial optimisation problem with important industrial applications. Although extensively studied, quantum solutions for instances beyond a dozen locations remain uncommon. We present a hybrid penalty-free Variational Quantum Algorithm (VQA) that significantly reduces both the qubit count and the runtime through a compact penalty-free formulation scaling as $\mathcal{O}(nlog_2 n)$ qubits compared to conventional formulations with $\mathcal{O}(n^2)$ scaling.
The VQA achieves solution qualities exceeding 90\% on Rigetti quantum hardware for instances of up to 17 locations. To the best of our knowledge, this represents the largest TSP instance solved on gate-based hardware. We systematically evaluate multiple encoding and optimisation strategies, and show that warm-start initialisation improves solution quality, while runtime is reduced by nearly two orders of magnitude through the  use of gradient-free optimisation methods and cost-function caching. We additionally introduce a novel VQA inspired classical machine learning model (VQAi-CML) and benchmark the VQA against VQAi-CML, a greedy heuristic, a Monte Carlo baseline, and a strong classical optimisation baseline. Across the networks evaluated, the VQA consistently achieves superior solution quality relative to VQAi-CML, greedy heuristic, and Monte Carlo baselines, but not the classical solver

\end{abstract}

\section{Introduction}

The well-known Travelling Salesman Problem (TSP) is an important NP-Hard combinatorial optimisation problem, with industrial use cases including last-mile delivery \cite{shi_exact_2026} and warehousing \cite{bock_survey_2025}.  In TSP, a salesman must visit each network location, and return to the start location, whilst minimising the cycle length.  There are powerful classical solvers \cite{ortools_routing,Concorde}, and classical algorithms as diverse as clustering routing \cite{dutta_heuristic_2025} and discrete artificial ant \cite{changdar_solving_2026} and bee \cite{li_discrete_2024} colony algorithms.  Recent classical implementations have solved networks with up to 25 million locations \cite{applegate_chained_2003}.  TSP has been studied on quantum annealing devices \cite{ciacco_quantum_2026,jain_solving_2021,bonomi_Quantum_2022,liu_quantum_2024}, and with the quantum circuit-model paradigm using variants of the hybrid Quantum Approximate Optimisation Algorithm (QAOA) algorithm \cite{hadfield_quantum_2018,farhi_Quantum_2014,bouchmal_quantum_2025,ruan_Quantum_2020,spyridis_variational_2023,garhofer_direct_2024}. Both quantum annealing and QAOA often use a Quadratic Unconstrained Binary Optimisation (QUBO) formulation \cite{ceselli_good_2023,koch_state_2025,lucas_ising_2014}. Recent formulations use fewer qubits than QUBOs \cite{vargas-calderon_many-qudit_2021, bentley_Quantum_2022,ramezani_reducing_2024,bako_prog-qaoa_2025, romero_bias-field_2025, zheng_constrained_2025}, culminating in penalty-free formulations \cite{goldsmith_beyond_2024, schnaus_efficient_2024,bourreau_indirect_2023}.  Despite this work, descriptions of quantum solutions of TSP networks of more than a dozen locations in the literature are rare, and we only found two studies of circuit model simulations of networks with eight locations, neither of which achieved greater than 99.7\% solution quality \cite{bai_quantum_2025, bourreau_indirect_2023}.

QUBO and Higher-order Unconstrained Binary Optimisation (HUBO) formulations consider, and then penalise, many invalid solutions, leading to a cost function that can be difficult to optimise.  Also, the often used parameter-shift optimiser can be slow, since a gradient evaluation is needed for each parameter, and this can take many shots.

This work studies simulations of a penalty-free hybrid circuit-model Variational Quantum Algorithm (VQA) and explicitly targets known bottlenecks such as qubit scaling, constraint handling, and embedding overhead that currently prevent quantum optimisation methods from engaging with realistic problem instances.  These bottlenecks are also relevant to a wide range of optimisation problems.  The model qubit count scales as $\mathcal{O}(nlog_2(n))$, rather than the QUBO scaling of $\mathcal{O}(n^2)$, where there are $n$ locations in the network.  Because of the favourable scaling, only 49 qubits are required to solve a network of 17 locations, whereas the QUBO formulation would require at least 256 qubits.  Computational time is reduced from over nine minutes to seven seconds for eight-location networks by deploying gradient-free optimisers and by caching cost function evaluations.

Although the individual quantum model components used, such as standard gates, circuits, and optimisers, are well known, our contribution lies in their integration into a penalty-free variational non-factorial formulation that fundamentally alters how feasibility constraints are enforced in TSP-like problems. To our knowledge, this represents a novel algorithmic construction rather than an application of an existing theoretical framework.

Whilst alternative quantum algorithmic frameworks such as the Quantum Singular Value Transformation (QSVT) \cite{gilyen_quantum_2019} have proven to be powerful for linear-algebraic problems, Hamiltonian simulation, and spectral transformations, they are less naturally aligned with combinatorial optimisation tasks such as the Travelling Salesman Problem. Applying QSVT would require efficient block-encoding of the TSP cost operator and a mechanism for extracting discrete tours from processed quantum states, both of which are likely to incur significant circuit depth and ancilla overhead. In contrast, our VQA model provides a more direct interface to discrete optimisation, enabling explicit control over feasibility through a problem-inspired ansätze. Given the NISQ-era focus of this work, VQA represents a more suitable and practically accessible framework for the penalty-free formulation explored here.  

Although quantum advantage cannot be claimed, as the circuits used can be efficiently simulated classically using Matrix Product State (MPS) methods, and a strong classical solver achieves a better solution, this work is still valuable as the penalty‑free encoding of problem constraints, use of warm starts, and gradient free optimisation address fundamental algorithmic challenges relevant across a range of combinatorial optimisation problems.

The contributions of this work are the following:
\begin{enumerate}
\item A hybrid, penalty-free Variational Quantum Algorithm (VQA) is executed on the Rigetti Cepheus 1-108-Q super-conducting quantum device, achieving optimal or near-optimal solutions with solution qualities exceeding 90\% for TSP networks of up to 17 locations.  To our knowledge, this is the first successful execution of VQA of a 17-location TSP network on quantum circuit model hardware.  For networks of up to nine locations, 100\% perfect solution quality is achieved, comparing favourable with all other circuit model studies we have found.  Good solution qualities are found in simulations of networks of up to 48 locations.  Since shallow circuits are used, the VQA model is likely to be less prone to barren plateaus because there are fewer parameters, and more resistant to noise because of the reduced qubit count and the shallower circuits employed.  

\item A systematic assessment of encoding strategies for quantum TSP solutions is provided.  Qubit count is reduced using factorial and non-factorial penalty-free formulations.  Warm-start initialisation using a greedy classical heuristic and the use of Gray codes are found to improve solution quality.  In addition, we achieve runtime reductions from 37 seconds to 7 seconds (8-location network) through caching of classical cost function evaluations.  Although this study is focused on Noisy Intermediate Scale Quantum (NISQ) devices, efficient problem formulations will still be necessary in the Fault Application Scale Quantum (FASQ) era. 

\item We evaluate several approaches to VQA cost-function optimisation, including cost aggregation over the lowest-cost bit strings, theoretical analysis of cost-function differentiability, and comparison of gradient-estimation methods. We find that gradient-free approaches, particularly the Simultaneous Perturbation Stochastic Approximation (SPSA) and CMA-ES (Covariance Matrix Adaptation Evolution Strategy) achieve comparable optimisation performance while reducing computational time from 217 seconds to 7 seconds (8-location network) relative to parameter-shift techniques. 

\item Additionally, we introduce Q-SPSA (Quantum-SPSA). Although the method is inspired by Nevergrad, we extend it by exposing and  tuning the most relevant hyper-parameters.  We further demonstrate how Q-SPSA's ability to reduce the number of classical cost-function evaluations per iteration can be leveraged in quantum optimization settings, where each evaluation requires a large number of shots and consequently incurs significant cost.

\item We develop a novel VQA-inspired classical machine learning model (VQAi-CML), in which the bit-string output distribution is governed by classical parameters rather than by quantum gate rotations (Section \ref{ml}).

\item A classical Monte Carlo benchmark, which samples from the same number of bit strings as the VQA or VQAi-CML models, is introduced alongside a greedy heuristic. To assess performance against a strong classical optimisation baseline, we compare solution quality with the Google OR-Tools Routing solver \cite{ortools_routing}.

\end{enumerate}

The remainder of this document is organised as follows.  Section \ref{Related_work}, describes how the TSP is solved using quantum annealing, the quantum circuit model, and bosonic sampling devices, and presents an evaluation of the limitations of existing methods.  Section \ref{Methods} describes our Methods, including the VQA model (Section \ref{var}), problem formulations (Section \ref{pre}), VQAi-CML (Section \ref{ml}), optimisation of the cost function (Section \ref{optimisation}) and an evaluation of runtime dependency on network size  (Section \ref{timing}).  Section \ref{Results} presents the results.  Section \ref{Conclusion} concludes the paper and discusses future work.  Parameter settings are placed in Appendix \ref{parameter_settings}, detailed results in Appendix \ref{appendix_results}, and figures in Appendix \ref{appendix_figures}.

\section{Solving the Travelling Salesman problem with a quantum device}
\label{Related_work}

The aim of the Travelling Salesman Problem (TSP) is to find the shortest Hamiltonian cycle that visits each of the $n$ network nodes once. The network can be represented as a graph $G=(V, E)$, with each node represented as one of the $V$ vertices in the graph, and the connections between the nodes represented as  $E$ edges.  Each edge $uv \in E $ between the vertices $\{u,v\}\in V$ has a distance $D_{u,v}$.  The problem is equivalent to finding a permutation $\pi$ of the $n$ nodes that minimises the distance, or cost, $C(\pi)$ shown in Equation \ref{eq:1}, noting that $\pi(n) = \pi(0)$ indicates that the starting location is revisited at the end of the cycle.  In this work, both vertices and nodes are used interchangeably to denote a location.   

\begin{equation} 
\label{eq:1} 
C(\pi) = \sum_{i=0}^{n-1}D_{\pi(i), \pi(i+1)} 
\end{equation}

\subsection{Quadratic Unconstrained Binary Optimisation}

Lucas \cite{lucas_ising_2014} formulates TSP as a Quadratic Unconstrained Binary Optimisation (QUBO) scaling as $\mathcal{O}(n^2)$.  In a QUBO, the problem Hamiltonian $H_{P}$ to be minimised is written as a function of binary variables $x_i$ of no higher than quadratic order, as shown in Equation \ref{eq:combo}: 

\begin{equation} \label{eq:combo} 
H_{P} = \sum_{i,j} Q_{i,j} x_i x_j
\end{equation}

The problem Hamiltonian is the sum of the objective function to be minimised, and a penalty Hamiltonian to represent constraints, multiplied by arbitrary Lagrangian multipliers to ensure that it is not energetically favourable to break the constraints.

The quadratic problem formulation in Equation \ref{eq:combo} can be mapped to an adiabatic quantum annealing device.  Networks of up to a dozen locations have been solved with the D-Wave quantum annealer \cite{ciacco_quantum_2026, jain_solving_2021,bonomi_Quantum_2022}, and networks of up to 48 locations where some computation is offloaded to classical devices. \cite{bonomi_Quantum_2022,liu_quantum_2024}. 

In the QAOA hybrid \cite{farhi_Quantum_2014,bouchmal_quantum_2025, ruan_Quantum_2020, spyridis_variational_2023, garhofer_direct_2024}, the time evolution of the adiabatic Hamiltonian is Trotterised using a parametrised quantum circuit with alternating layers, and the parameters are optimised classically.  The problem Hamiltonian is normally the same QUBO formulation. QAOA is an example of a Variational Quantum Algorithm (VQA) \cite{qi_variational_2024} where the parameters are classically optimised in a feedback loop. In the  circuit model paradigm, the parameters are rotations \cite{amaro_case_2022, patti_variational_2022}, whereas boson samplers parameterise the beam splitter and/or phase shift angles \cite{slysz_solving_2026,goldsmith_beyond_2024}.  

\subsection{Non-QUBO formulations}
The QUBO formulation above uses a One-Hot encoding scheme, which is inefficient since only a few binary output strings are valid solutions.  Recent proposed formulations use binary variables more efficiently, which could allow devices to solve larger problems \cite{vargas-calderon_many-qudit_2021, bentley_Quantum_2022}.   Higher-order Unconstrained Binary Optimisation (HUBO) formulations \cite{bako_prog-qaoa_2025, ramezani_reducing_2024, romero_bias-field_2025, zheng_constrained_2025} have reduced the number of qubits required $\mathcal{O}(nlog_{2}n)$, but at the expense of more two-qubit gates.

\subsection{Penalty-free formulations}
\label{beyond}

Recently, quantum penalty-free formulations have been implemented in VQAs \cite{goldsmith_beyond_2024, schnaus_efficient_2024, bourreau_indirect_2023}, with the number of qubits required scaling as $\mathcal{O}(nlog_{2}n)$ where $n$ is the number of locations, and, by design, all bit strings are mapped to valid cycles, avoiding penalty terms.  \cite{goldsmith_beyond_2024} introduces a \emph{non-factorial} penalty-free formulation as described in Section \ref{non-factorial} and simulate this formulation for networks of up to 48 locations on a quantum boson sampler.

In contrast, \cite{schnaus_efficient_2024} and \cite{bourreau_indirect_2023} implement a \emph{factorial formulation} built on earlier work  \cite{sedgewick_permutation_1977}, where the possible $n!$ permutations of the $n$ locations are enumerated, with each permutation representing a valid cycle, as described in Section \ref{factorial}. The factorial formulation is simulated in \cite{schnaus_efficient_2024} on a network of six locations using a Variational Quantum Eigensolver algorithm (VQE) and in \cite{bourreau_indirect_2023} on a network of eight locations using Indirect Quantum Approximate Optimisation Algorithm (IQAOA), a variant of QAOA.  Neither simulation achieved 100\% solution quality. 

The use of problem-inspired ansätze to encode feasibility constraints directly within the circuit, thereby avoiding penalty terms in the cost Hamiltonian, is well established in the QAOA literature through the Quantum Alternating Operator Ansatz framework of \cite{hadfield_quantum_2018}. In the context of the TSP, swap-based and permutation-preserving mixers are proposed to ensure that quantum evolution remains within the feasible tour subspace throughout the optimisation process.  However, this approach may be challenging to implement on NISQ devices as it scales as $\mathcal{O}(n^2$) and the mixers, with unitary row swaps \emph{"contain $2n$-local interactions and are non-trivial to implement"}.

\subsection{Other quantum approaches}
Recent studies highlight the range of possible quantum algorithms available.  \cite{bai_quantum_2025} generate the uniform superposition state of all n-length Hamiltonian and claim to use very few qubits, achieving an accuracy of 99.7\% for eight locations.  \cite{das_quantum-inspired_2023} uses a Quantum-inspired Ant Colony Optimisation (Qi-ACO) to solve a four-dimensional travelling salesman problem, and \cite{li_quantum_2024} includes quantum rotation gates and qubits into a traditional ant colony optimisation, finding that efficiency and convergence speed were enhanced.   \cite{li_solving_2024} uses a quantum self-attentive neural network integrated with a variational quantum circuit, trained using deep reinforcement learning algorithms, and reports reduced training parameters and dataset size while achieving superior optimisation results compared to classical methods.  Four location networks were solved by Sato et al. in 2025 \cite{sato_two-step_2025} simulating a two step algorithm using Grover’s search algorithm, and by Baldazi et al. in 2026 \cite{baldazzi_resource-efficient_2026} entangling registers representing path arrival and departure locations on a photonic device.

Although parameter-shift optimisation remains a widely used gradient-based technique, \cite{skalli_model-free_2025} studied gradient-free techniques and observed that SPSA offers favourable optimisation performance for circuits with relatively few parameters, whereas CMA-ES and PEPG (Parameter-Exploring Policy Gradients) become more effective as the dimensionality of the parameter space increases.

\subsection{Solving TSP with classical neural networks}

There is a long history of solving TSP with classical neural networks \cite{sui_survey_2024}.  Although Hopfield networks were originally proposed in 1982 for associative memory \cite{hopfield1982neural}, in 1985 Hopfield and Tank \cite{hopfield_neural_1985} used a Hopfield network instantiated on an analogue electrical circuit to solve TSP on what was effectively a QUBO formulation and reported simulations of network solutions of up to ten locations on an early digital computer.  \cite{feng_using_2000} built on this work to find valid solutions to a 51-location TSP, on average only 15\% longer than the optimal cycle. \cite{siqueira_recurrent_2008} used a recurrent neural network with a \emph{winner take all} algorithm to solve networks of up to 532 locations.  \cite{la_maire_comparison_2012} found that in networks of up to 80 locations Integer Programming and Kohonen Neural Networks resulted in close to optimal solutions, whereas the Hopfield Neural Networks performed less well.  

\subsection {Warm starts for quantum algorithms}
\label{warm_int}

Barren plateaus \cite{larocca_barren_2025} are a significant issue for VQAs.  \cite{cerezo_does_2024} argue that when a problem exhibits a barren plateau, its loss function becomes \emph{"on average, exponentially concentrated with the system size"}, making optimisation difficult, and that, if structures exist that avoid barren plateaus, the quantum device can be simulated classically. Rather than using a \emph{cold start}, such as $\ket{0}^{\otimes n}$ or $\ket{+}^{\otimes n}$, a \emph{warm start}, where an initial state $\ket{\psi_{\mathrm{init}}}$ is derived from an approximate classical solution, may help avoid vanishing gradients with barren plateaus, as investigated in \cite{tate_theoretical_2025, sauvage_flip_2021,polina_hot-start_2021,recio-armengol_train_2025}.

\section{Methods}
This section describes the two Variational Algorithms implemented (Section \ref{var}): the Variational Quantum Algorithm (VQA) (Section \ref{qc}), and the novel classical VQAi-CML model (Section \ref{ml}). Two penalty-free formulations that map from bit strings to valid cycles are contrasted: the factorial formulation (Section \ref{factorial}) and the non-factorial problem formulations (Section \ref{non-factorial}). Gray encoding (Section \ref{gray}) and caching (Section \ref{cache}) are discussed.  In the optimisation section (Section \ref{optimisation}) averaging by slicing (Section \ref{av}) is explained, and the differentiability of the cost function is demonstrated (Section \ref{diff}). Parameter shift (Section \ref{parameter_shift}) is compared to gradient free optimisers such as SPSA and Q-SPSA (Section \ref{SPSA}) and CMA-ES (Section \ref{grad_free}). A warm start protocol is described (Section \ref{warm}).  An evaluation of run-time dependency on model size is highlighted. (Section \ref{timing}).  
\label{Methods}
\subsection{Variational Algorithms}
\label{var}

\begin{figure}[H]
    \centering
    \includegraphics[width=1.0\linewidth]{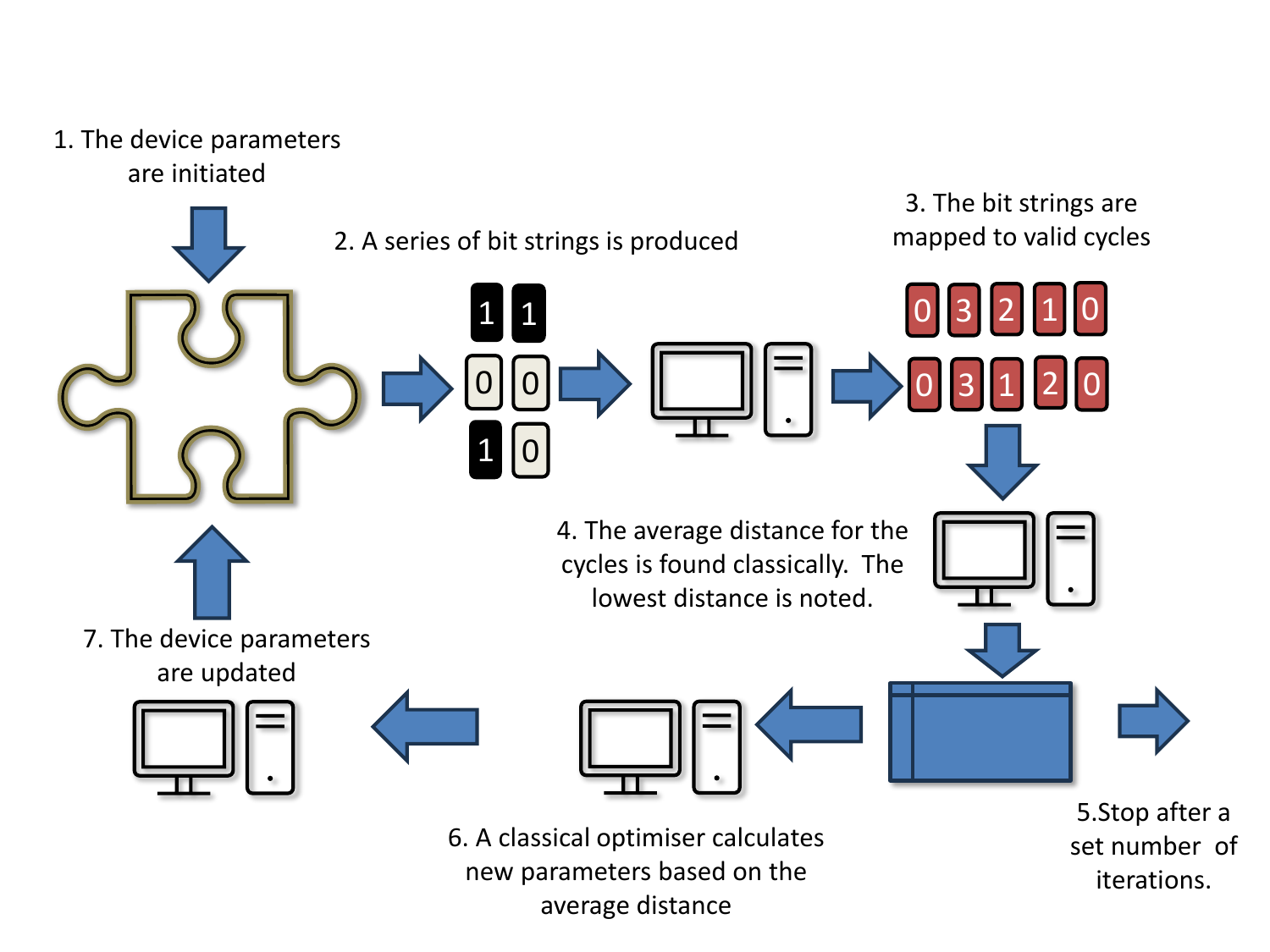}
    \caption{An overview of a Variational Algorithm, showing how either a classical or quantum device sample bit strings, which are mapped to cycles, and an average distance evaluated. A classical optimiser updates the device parameters in a feedback loop.}
    \label{fig:overview}
\end{figure}

A Variational Algorithm, as shown in Figure \ref{fig:overview}, is studied.  A device, either quantum or classical, produces samples of bit strings, with a probability distribution dependent on the values of adjustable parameters.  The parameters are optimised classically in a feedback loop.  In the quantum VQA model gate rotations are parametrised (Section \ref{qc}), whereas in the classical VQAi-CML model (Section \ref{ml}) the parameters are model weights.  The steps followed are as follows:
\begin{enumerate}
    \item  The device parameters are initialised with constant values or by a warm start, as explained in Section \ref{warm}.
    \item Bit strings are sampled from the device.
    \item Bit strings are mapped to routes using either the non-factorial formulation (Section \ref{non-factorial}) or the factorial formulation  (Section \ref{factorial}), which constrain each bit string interpretation to a valid route, avoiding penalty terms.  In these formulations, bit strings can be interpreted as integers using either a binary string or a Gray code (Section \ref{gray}).
    \item The distance for each cycle is found classically, and an average distance for the bit strings sampled is calculated.  The average used is either a simple average or a fraction of the lowest cost bit strings (Section \ref{av}).
    \item  The lowest distance found to date is recorded and, after a set number of iterations, the loop is terminated, and the results are output.
    \item A classical optimiser uses gradient descent, as described in Section \ref{optimisation}, to calculate new parameter values. 
    \item The device parameters are updated.
    \item The cycle is re-executed from Step 2.
\end{enumerate}

\FloatBarrier

\subsubsection{Quantum circuits in VQA}
\label{qc}

Six quantum circuits were tested in the VQA.  These circuits are chosen because they are shallow and only require  limited gate connectivity: suitable for NISQ-era devices.  Each $R_X$, $R_Y$, $R_{XX}$ and $R_{ZZ}$ gate has a rotation angle that can act as a tunable parameter.  Circuits 1, 2a, 2b and 3 introduce entanglement using a qubit ring which optimises Rigetti Cepheus connectivity since it can be deployed without extra swap gates. Circuits 4 and 5 are entanglement-free. Circuit 3 is from the well-studied IQP family, whereas Circuit 1, 2a and 2b are of our own design.

\begin{figure}[H]
    \centering
    \begin{subfigure}[b]{0.45\textwidth}
        \centering
        \includegraphics[width=1\linewidth]{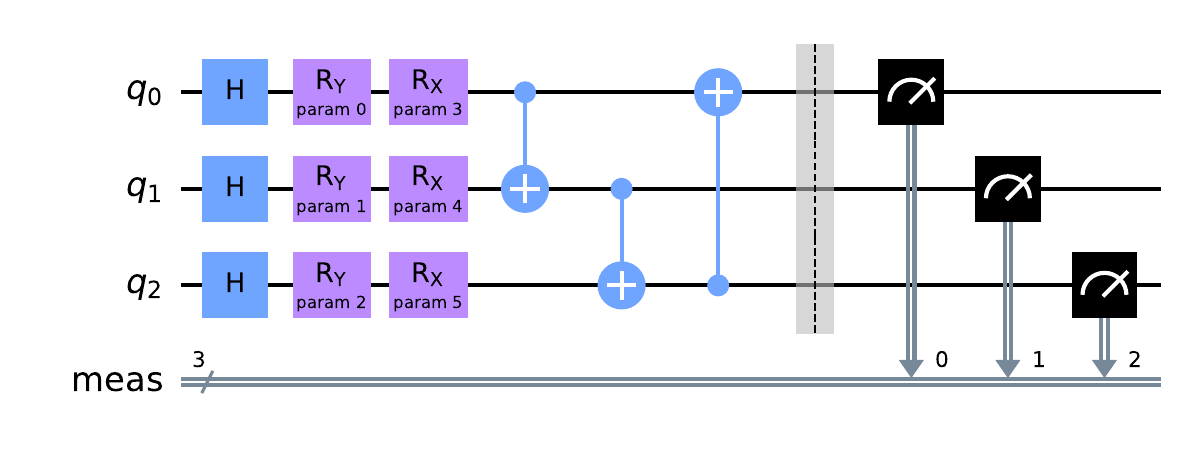}
        \caption{\textbf{Circuit 1}: RY, RX and entangling CX}
        \label{fig:model1}
    \end{subfigure}
    \hfill
    \begin{subfigure}[b]{0.45\textwidth}
        \centering
        \includegraphics[width=1\linewidth]{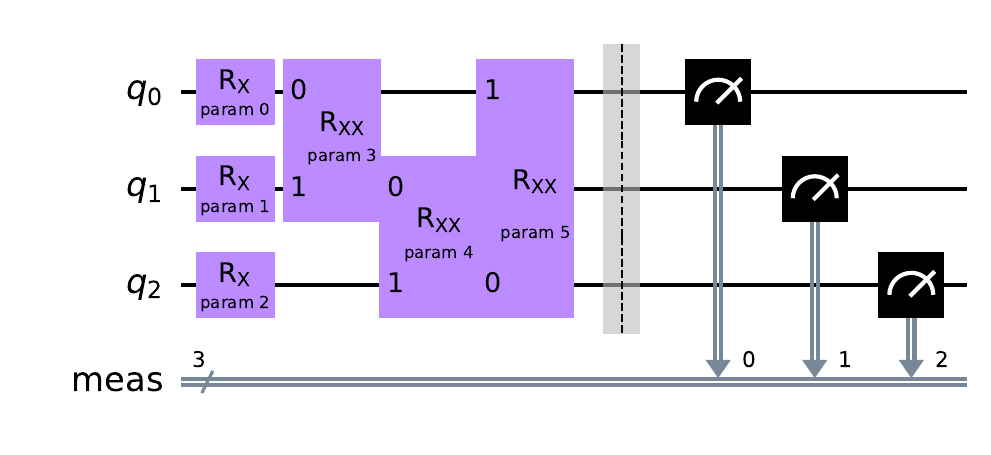}
        \caption{\textbf{Circuit 2a}: RX and entangling RXX}
        \label{fig:model2a}
    \end{subfigure}

    \begin{subfigure}[b]{1.0\textwidth}
        \centering
        \includegraphics[width=1.0\linewidth]{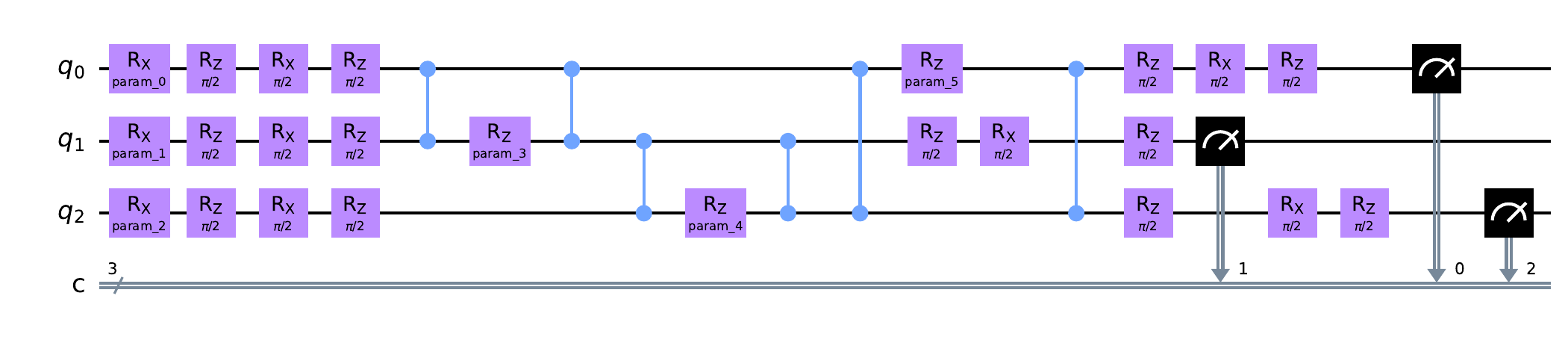}
        \caption{\textbf{Circuit 2b}: Circuit 2a transpiled to the Rigetti Cepheus 1-108-Q gateset}
        \label{fig:model2b}
    \end{subfigure}

    \begin{subfigure}[b]{1.0\textwidth}
        \centering
        \includegraphics[width=1.0\linewidth]{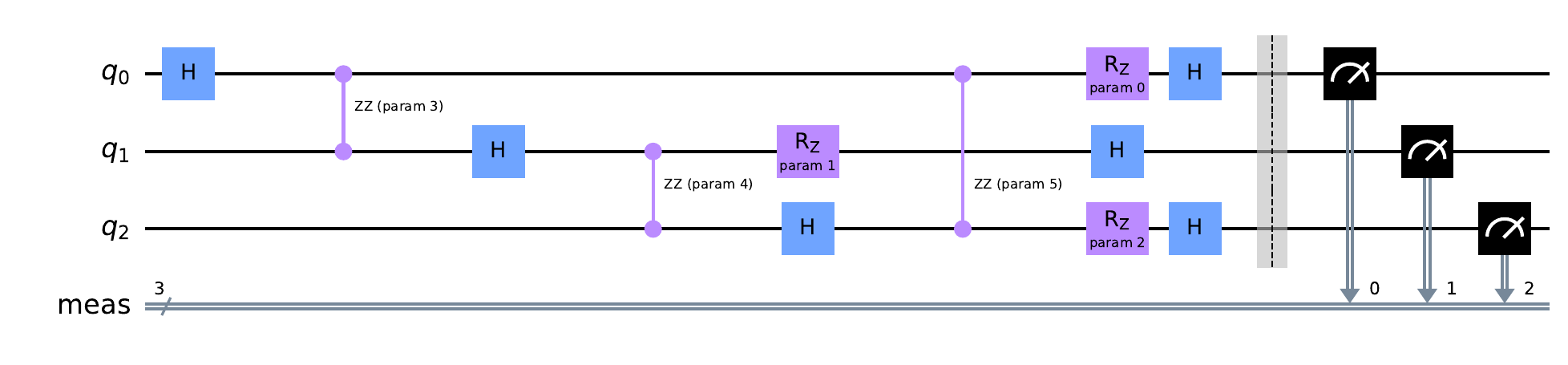}
        \caption{\textbf{Circuit 3}: IQP}
        \label{fig:model3}
    \end{subfigure}

    \begin{subfigure}[b]{0.45\textwidth}
        \centering
        \includegraphics[width=0.8\linewidth]{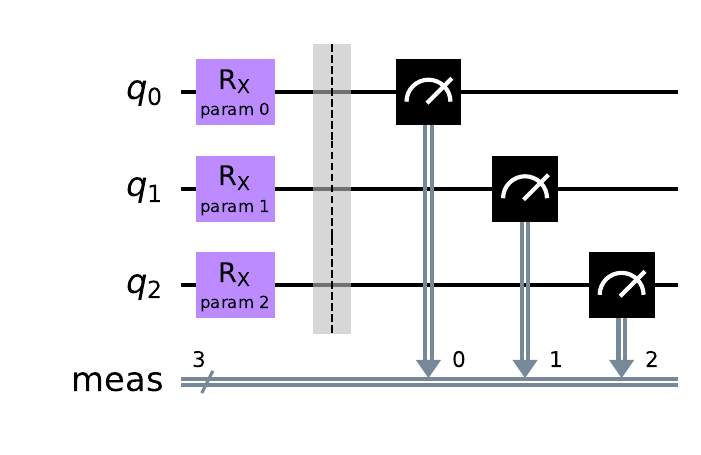}
        \caption{\textbf{Circuit 4}: RX and no entangling gates}
        \label{fig:model4}
    \end{subfigure}
    \hfill
    \begin{subfigure}[b]{0.45\textwidth}
        \centering
        \includegraphics[width=0.8\linewidth]{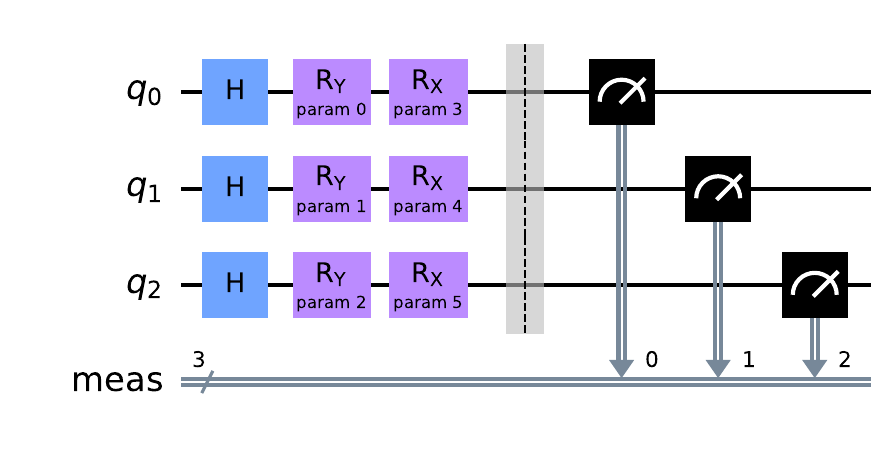}
        \caption{\textbf{Circuit 5}: H, RY, RX and no entangling gates}
        \label{fig:model6}
    \end{subfigure}
    \end{figure}

\begin{itemize}
    \item \textbf{Circuit 1} has a Hadamard gate ($H$), a gate $R_X$ and $R_Y$ for each qubit, and a $CX$ entangling gate linking sequential qubits (Figure \ref{fig:model1}).  
    \item \textbf{Circuit 2a} has an $R_X$ gate for each qubit and entangling $R_{XX}$ gates connecting sequential qubits (Figure \ref{fig:model2a}) and supports a warm start.  
    \item \textbf{Circuit 2b} is the same as Circuit 2a, except that it uses the Rigetti Cepheus 1-108-Q device gate set : $R_X(\pi)$, $R_X(\frac{\pi}{2}$), $R_X(-\frac{\pi}{2}$), $R_Z$ and $CZ$ (Figure \ref{fig:model2b}). 
    \item \textbf{Circuit 3} is an IQP circuit with 2-local $ZZ$ sandwiched between $H$ gates at the start and end of the circuit. (Figure \ref{fig:model3}). 
    \item \textbf{Circuit 4} is entanglement-free and uses only $R_X$ gates and supports a warm start. (Figure \ref{fig:model4}).
    \item \textbf{Circuit 5} is also entanglement-free, and uses $H$, $R_Y$, and $R_X$ gates (Figure \ref{fig:model6}).

\end{itemize} 

Circuits 4 and 5 are known to be classically simulable, since there are no entangling gates.  Circuit 1 includes a final layer of nearest-neighbour $CX$ gates applied immediately before measurement which are equivalent to an invertible linear transformation of the measured bit string, so is also classically simulable.  Circuit 2 and 3 were found to classically simulated for circuits of up to 219 qubits using the Qiskit Matrix Product State simulator.  There are at most two bits of entanglement across any cut, so the circuit can be represented by an MPS of bond dimension 4.

\FloatBarrier

\subsection{Problem formulation}
\label{pre}

This section describes how a bit string is converted into a valid route, for which the distance can be evaluated.  Both a non-factorial and factorial penalty free formulations are described, which do not require specialised mixers, as deployed by Hadfield \cite{hadfield_quantum_2018}, and scale more favourably $\mathcal{O}(n^2)$ qubits rather than $\mathcal{O}(nlog_2 n)$.  Gray encoding is studied and the computational time is reduced by caching the classical distance evaluations.

\subsubsection{Non-factorial formulations without penalty terms}
\label{non-factorial}
\begin{figure}[H]
    \centering
    \includegraphics[width=0.8\linewidth]{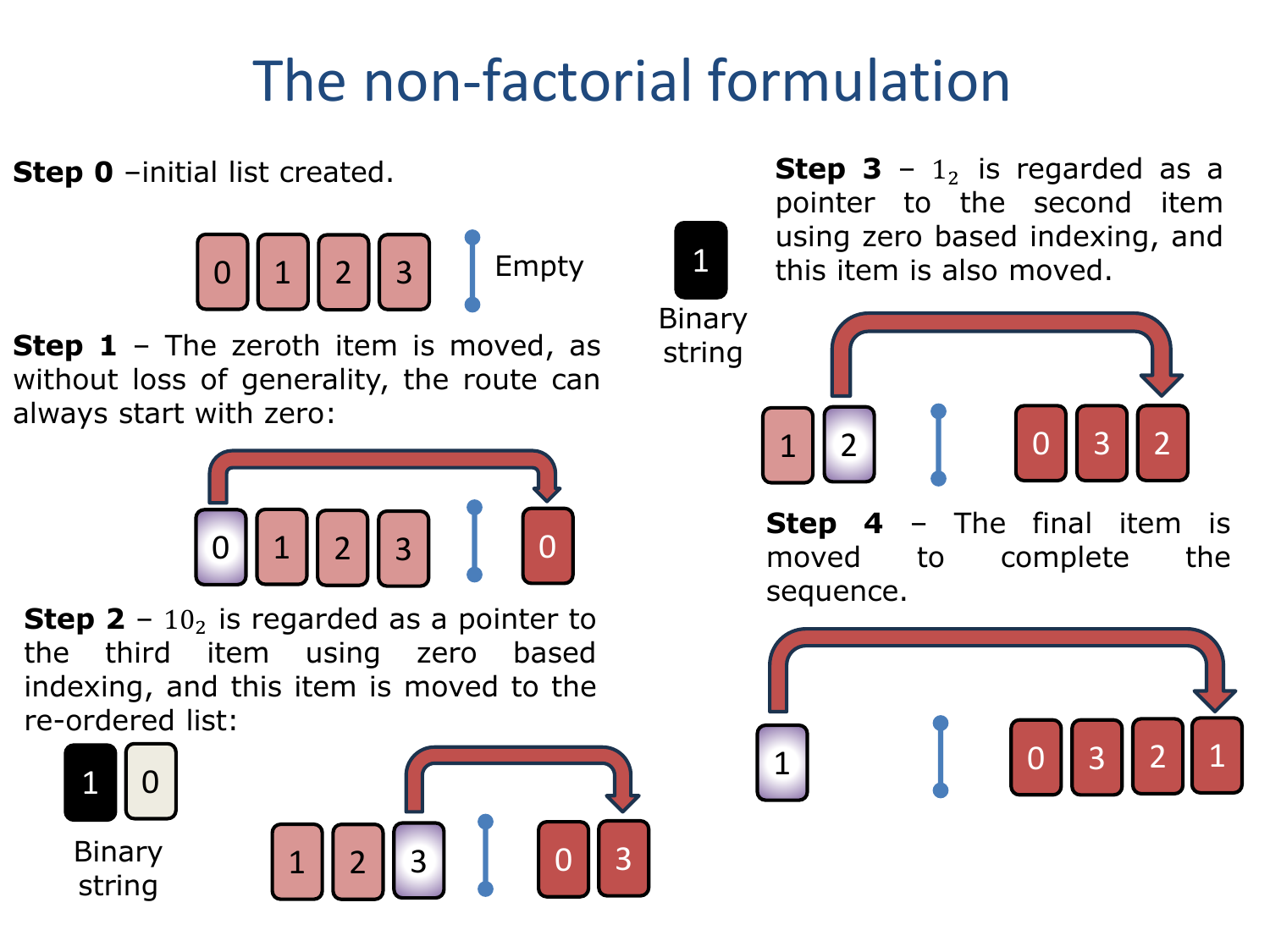}
    \caption{\textbf{Algorithm 1}: Diagram showing how the bit string $10_2 1_2$ is used to construct a cycle by using the bit string to point to the next item of the ordered list to be moved}
    \label{fig:form1}
\end{figure}

A non-factorial formulation without penalty terms is implemented, following \cite{goldsmith_beyond_2024}.  Figure \ref{fig:form1} and Algorithm 1 show how location references are moved from an initial sequential list to a reordered list until the initial list is empty and the reordered list contains a valid cycle. This is done by splitting the bit string sampled from the quantum device into smaller strings, which are interpreted as binary encoding of indices that point to the next item of the ordered list to be moved.  The use of modulo arithmetic prevents errors where the index is too high to reference a valid element of the ordered list. The length of the required bit string $l_b$ is calculated as shown in Equation \ref{eq:len2}:

\begin{equation} \label{eq:len2} 
    l_b = \sum _{i=1} ^{n-1} \left\lceil \log_2(i) \right\rceil 
\end{equation}

\begin{algorithm}[htbp]
\caption{Non-factorial formulation from \cite{goldsmith_beyond_2024}.  This algorithm receives a bit string $b$ of length $l_b$ as defined in Equation \ref{eq:len2}, and the number of locations $n$ as an input. The output is a permutation of the numbers $\{0, \dots, n-1\}$, which can be interpreted as a cycle in a TSP with $n$ locations.  This algorithm uses zero-based indexing} 
\label{algo:1} 
\begin{algorithmic}[1] 
\Require Bit string $b$ of length $l_b$, number of locations $n$ 
\Ensure A valid permutation $s$ of the locations $\{0,\ldots,n-1\}$
\State $cycle \gets (1,\dots,n-1)$ \State $s \gets (0)$ 
\State $i \gets n-1$ 
\While{$i > 1$} 
\State Extract the next $\lceil \log_2(i) \rceil$ bits from $b$ as $b_{\text{temp}}$ 
\State $j \gets \mathrm{int}(b_{\text{temp}})$ 
\State $k \gets j \bmod i$ 
\State Append $cycle_k$ to $s$ 
\State Remove $cycle_k$ from $cycle$ 
\State $i \gets i-1$ 
\EndWhile 
\State Append the remaining element of $cycle$ to $s$ 
\State \Return $s$ 
\end{algorithmic} 
\end{algorithm}

\subsubsection{The factorial formulation without penalty terms}
\label{factorial}

In contrast to the \emph{non-factorial formulation}, \cite{schnaus_efficient_2024} and \cite{bourreau_indirect_2023} use a \emph{factorial formulation} to enumerate the possible $n!$ permutations of the $n$ locations, noting that each permutation represents a valid cycle.  By generalising standard positional number systems, such as decimal, an index $x$ that identifies a unique permutation can be written as shown in Equation \ref{facts}:

\begin{equation} 
    \label{facts}
    x = \sum_{i=1}^{i=n} a_i i!
\end{equation}
where $a_i \leq i!$.  Algorithm \ref{algo:2} shows how the $a_i$ are calculated for a binary string generated by a quantum device, and are used to identify items to be moved from an initial sequential list to a re-ordered list.

\begin{algorithm}[htbp] \caption{Factorial formulation based on  \cite{schnaus_efficient_2024}. This algorithm receives a measured state $x$, which is interpreted as a binary number, and the total number of locations $n$ as an input. The output is a permutation of the numbers $\{1, \dots, n\}$, which can be interpreted as a cycle in a TSP with $n$ nodes. The formulation in the original paper has been slightly modified.} 
\label{algo:2} 
\begin{algorithmic}[1] 
\Require Index $x$, number of locations $n$ 
\Ensure A valid permutation $s$ of the locations $\{0,\ldots,n-1\}$
\State $f \gets n!$ 
\State $y \gets x \bmod f$ 
\State $nodes \gets (1,\dots,n)$ 
\State $s \gets ()$ 
\State $i \gets 0$ 
\While{$i < n$} 
\State $f \gets \frac{f}{n-i}$ 
\State $k \gets \left\lfloor \frac{y}{f} \right\rfloor$ 
\State Append $nodes_k$ to $s$ 
\State Delete $nodes_k$ from $nodes$ 
\State $y \gets y - kf$ 
\State $i \gets i + 1$ 
\EndWhile 
\State \Return $s$
\end{algorithmic} 
\end{algorithm}

\FloatBarrier

\subsubsection{Gray encoding}
\label{gray}

To interpret bit strings as a numerical index, two approaches are compared: a standard binary coding, where the bit string is regarded as the binary representation of a number; and Gray encoding, a bijective map between bit strings and numbers, designed so that as the number is increased by one, only one bit of the string changes.

\subsubsection{Caching}
\label{cache}
The classical mapping of a bit string to a cycle and the calculation of the distance for that cycle may be performed many times for a particular bit string.  To reduce computational time, the results for each bit string are cached in memory. A custom implementation is required because the Last Recently Used (LRU) cache provided by the \texttt{@lru\_cache} decorator in the Python \texttt{functools} package assumes that the key of the cache table is a string, whereas in our code, the key is a list output by Qiskit.

\subsection{VQA-inspired Classical machine learning model (VQAi-CML)}
\label{ml}

\begin{figure}[htbp]
    \centering
    \includegraphics[width=1.0\linewidth]{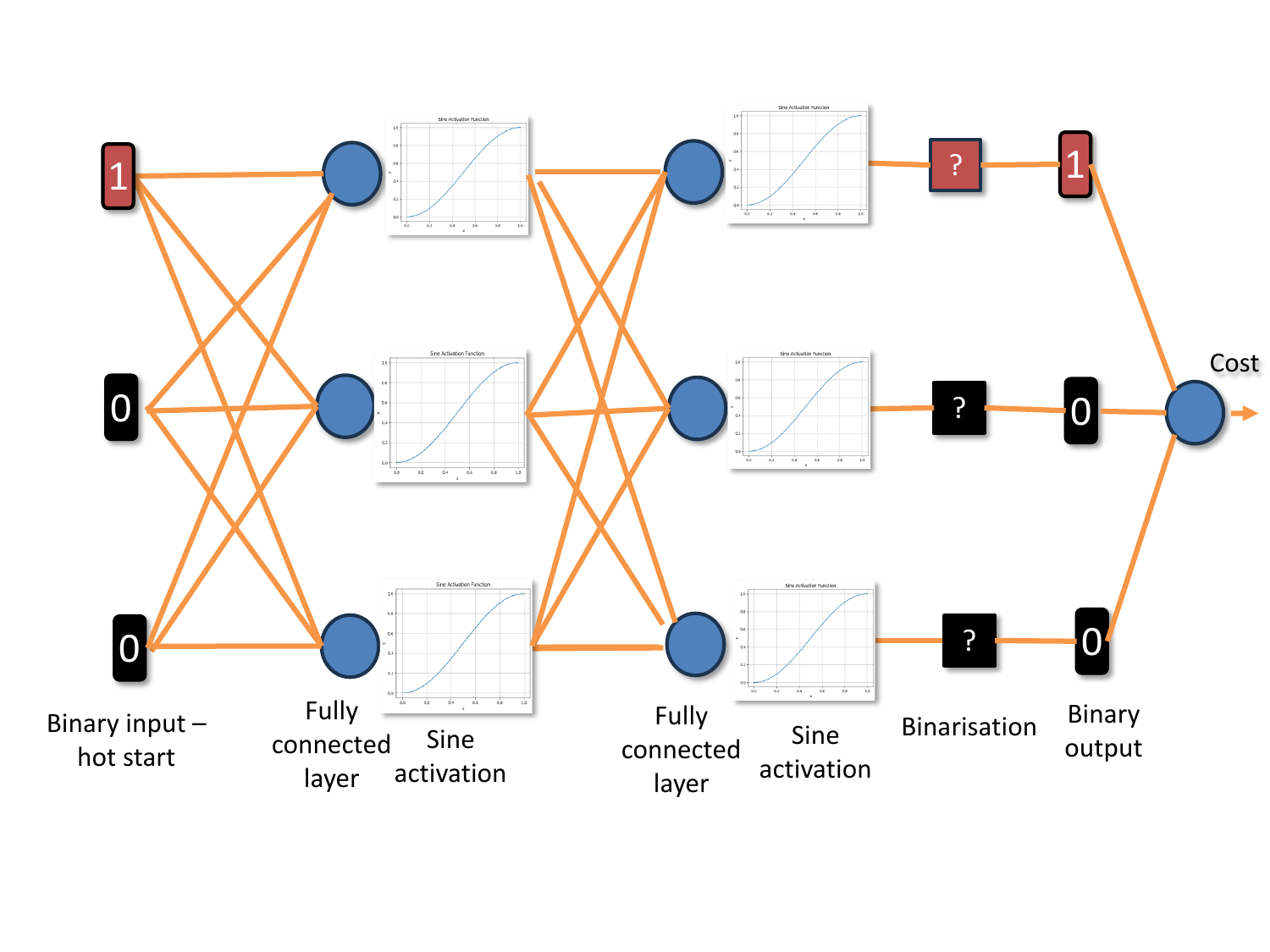}
    \caption{VQAi-CML with two fully connected layers, Sine activation, binarisation, and objective function evaluation}
    \label{fig:ML}
\end{figure}

A novel VQA-inspired Classical machine learning model (VQAi-CML) is shown in Figure \ref{fig:ML}.  The VQAi-CML model has the same number of inputs and outputs as the VQA model, uses fully connected layers, a sine activation function inspired by the quantum model, and produces a binary output by comparing the output of each activation layer to a random number.

\subsubsection{Input and fully connected layer}
\label{ml_warm}

The input data are structured as a $N_s \times q$ PyTorch tensor, comprising $N_s$ identical vectors of dimension $q$, where $q$ is the number of binary variables required to model the $n$ locations. The number of input vectors $N_s$ is equivalent to the number of shots in the quantum model, or an ML mini-batch, and is a tunable hyper-parameter.  The number of binary variables, q, is equivalent to the number of qubits in the quantum algorithm.  The circuit is shown in Figure \ref{fig:ML}, where $q = 3$.  With a warm start, each element of the input is initialised by setting all biases and weights to zero, except for the weights for the diagonal elements of the adjacency matrix for the fully connected layer, which are set to 1, so that the input bit strings propagate to the output of the model. If there is no warm start, the input is either set to 0, or to 0.5 and PyTorch automatically initialises the weights and biases. The input is fed into a fully connected layer with $q$ inputs and $q$ outputs.

\subsubsection{Activation function}
To introduce non-linearity, a Sine activation function mimics the trigonometric rotations of quantum devices.  

\subsubsection{Binarisation}
The output of the model is binarised depending on whether the input $x$ is greater than a random number $u$ of a uniform distribution in the interval $[0,1)$, as shown in Equation \ref{eq:binary_output}:

\begin{equation}
    \label{eq:binary_output}
    g(x) = 
    \begin{cases} 
    1, & \text{if } x > u \\
    0, & \text{otherwise}
    \end{cases}
\end{equation}

The model is coded in PyTorch \cite{paszke_pytorch_2019} and the gradients for the fully connected layer and the Sine activation are calculated automatically as standard.  The correct value for the binarisation gradient is $g(x) = x + (g(x) - x)$ and so $g(x) - x$ is detached from the computational graph, used to compute the gradients, ensuring that the gradient is correctly evaluated as the gradient of $x$.  

\subsubsection{Distance evaluation}

Similarly to the input vector, the data after binarisation is structured as a $N_s \times q$ tensor comprising $N_s$ output binary vectors of dimension $q$.  Let $\mathbf{x_i}$ denote the binary output vector $i^{th}$, where $\mathbf{x_i} \in \{0,1\}^q$.  $x_i^{(j)}$ refers to the $j^{th}$ bit of $\mathbf{x_i}$. Each of the binary output vectors $N_s$ is mapped to a valid cycle, and the distance for that cycle is evaluated using a classical  function module represented as a function $h : \{0,1\}^q \mapsto \mathbb{R}$.  The final output of the model at the end of each iteration is a simple average $\bar{d}$ of the distances calculated, as shown in Equation \ref{eq:av}, and described in Section \ref{av} below.

Bespoke coding is  needed to calculate and save the gradient in the forward pass and to retrieve the saved tensor in the backward pass.  The $j^{th}$ component of the  $i^{th}$ gradient vector is calculated by flipping the $j^{th}$ bit and evaluating the cost function before and after the bit-flip as shown in Equation \ref{eq:grad}, where $\oplus$ represents a bitwise XOR, and $\mathbf{e}^{(j)}$ represents the standard basis vector with 1 in the $j^{th}$ position and 0 elsewhere, noting that $2x_i^{(j)} - 1:\{0,1\} \mapsto \{-1,1\}$ is an adjustment that ensures the sign is correct.

\begin{equation}
    \frac{\partial h}{\partial x_i^{(j)}} = \frac{h(\mathbf{x_i}) - h(\mathbf{x_i} \oplus \mathbf{e}^{(j)})} {2x_i^{(j)} - 1}
    \label{eq:grad}
\end{equation}


\subsubsection{Gradient descent for VQAi-CML}
\label{SGD}

The standard PyTorch optimiser for gradient descent with SGD and Adam is compared.  In addition to the learning rate, the momentum and weight decay, an L2 penalty against weights, are treated as hyper-parameters.  Because back-propagation can be applied, gradient-free optimisers were not investigated.

\FloatBarrier

\subsection {Optimisation}
\label{optimisation}

\subsubsection{Averaging by slicing}
\label{av}

Each shot (bit string sampled) in an iteration could have a different cycle and a different distance.  The default average $\bar{d}$ over the $N_s$ shots in an iteration is shown in Equation \ref{eq:av}, recalling that $ h: \{0,1\}^n \mapsto \mathbb{R}$ is the function that assigns a scalar output to each bit string $\bm{x_i}$ by mapping the bit string to a valid cycle and finding the distance for that cycle.

\begin{equation}
    \label{eq:av}
    \bar{d} = \frac{1}{N_s} \sum_{i=1}^{N_s} h(\mathbf{x_i})
\end{equation}

Alternatively, the outputs $\{h(\mathbf{x_i})\}$ can be listed in ascending order of distance $[d_k]$ with only a fraction, or \emph{slice}, of the results used in the average, possibly reducing the sample noise.  The average $\bar{d_S}$ is calculated as shown in Equation \ref{eq:slice} where $N$ is the number of shots and $S$ is the slice fraction with $0 \le S \le 1$.

\begin{equation}
    \bar{d_S} = \frac{1}{S N_s} \sum_{k=1}^{S N_s} d_k
    \label{eq:slice}
\end{equation}

For example, if $S = 1$ the optimiser uses the simple average over all shots, if $S = 0.8$ the optimiser uses the average over the $80\%$ of the shots with the lowest distance.  Both the simple average and average by slicing are assessed.

\subsubsection{Differentiability of the cost function}
\label{diff}

It is known that in hybrid algorithms such as the Quantum Variational Eigensolver (QVE) and QAOA, the weighted sum of the measured expectation values can be minimised by tuning the parameters $\bm{\theta}$ \cite{mitarai_quantum_2018}.  However, with VQE and QAOA, the quantity being optimised is a quantum observable, whereas the VQA model studied optimises an output, written as $\mathcal{H}(\bm{\theta})$, obtained by mapping each bit string sampled to a valid cycle, and averaging the calculated distance over many samples to yield an estimate of the expected value, as defined in Equations \ref{eq:av} and \ref{eq:slice}. Therefore, it is necessary to demonstrate that the output of the VQA model is differentiable.

Our quantum circuits effect a unitary transformation $U(\bm{\theta})$ on an input $\ket{\psi_{\text{in}}}$.  Repeated measurement of the output wave function produces a series of classical bit strings $\bm{x_i} \in \{0,1\}^n$, each corresponding to an output wave function $\ket{x_i}$ with probability $p_i(\bm{\theta})$, given by Born's rule in Equation \ref{eq:prob}, where $p(\bm{\theta)}$ denotes the probability distribution from sampling the output bit strings:

\begin{equation}
    \label{eq:prob}
    p_i(\bm{\theta})
      = | \bra{x_i}\, U(\bm{\theta}) \ket{\psi_{\text{in}}}|^2
\end{equation}

 The expected output is defined by rearranging Equation \ref{eq:av} as Equation \ref{eq:output}:

\begin{equation}
    \label{eq:output}
    \hat {\mathcal{H}}(\bm{\theta}) := \bar{d}= \mathbb{E}_{\bm{x_i} \sim p(\bm{\theta)}} \sum_{\bm{x_i}} p_i(\bm{\theta)} \cdot h(\bm{x_i}).
\end{equation}

The function $\theta \mapsto p_i(\theta)$ is differentiable, since the VQA model uses parametrised quantum circuits with smooth gates \cite{mitarai_quantum_2018}.  $h(\bm{x_i})$ does not depend on $\bm{\theta}$, so the expected output $\bm{\theta}$ is also differentiable and its gradient is:

\begin{equation}
    \label{eq:chain}
    \frac{d \hat {\mathcal{H}}(\bm{\theta)}}{d\bm{\theta}} = \mathbb{E}_{\bm{x_i} \sim p(\bm{\theta)}} \sum_{\bm{x_i}} \frac{d p_i(\bm{\theta})}{d\bm{\theta}} \cdot h(\bm{x_i}).
\end{equation}

The expectation value estimated in Equation \ref{eq:output}, uses sampling using a finite number of shots ($N_S$) over a probability distribution ${p(\bm{\theta}})$.  Although this estimator is not strictly differentiable since it is based on finite samples, experience from quantum machine learning has shown that unbiased estimators of the gradient can be computed using methods such as the parameter-shift rule.  This generalisation might not apply to slicing (Equation \ref{eq:slice}), since only a subset of the sample is considered.  

This establishes that for the VQA model studied, the expected value of the output of the quantum device $\hat{\mathcal{H}}(\bm{\theta})$ can be treated as differentiable.  Now we know it is meaningful to discuss a gradient, and we compare different optimisation methods: parameter shift, and gradient methods such as SPSA, Q-SPSA and CMA-ES.

\subsubsection {Parameter Shift gradient estimation}
\label{parameter_shift}

A standard method for evaluating the gradient of quantum circuits is the parameter shift rule \cite{schuld_evaluating_2019}.   The $i^{th}$ component of the gradient $\bm{\nabla_t }(\bm{\theta_{t}})$ for the cost function ${\mathcal{H}(\bm{\theta})}$ in the $t^{th}$ iteration is shown in Equation \ref{eq:parameter_shift}, where $\bm{e_i}$ is a standard basis vector with all zeros except one at the $i^{th}$ position,  $s$ is a hyper-parameter with a default value of $0.5$ and $\langle \dots \rangle$ denotes an expectation value estimated by averaging over a number of shots. \textbf{Bold} is used to denote vectors such as $\bm{\nabla_t}$, $\bm{\theta_t}$ and $\bm{e_i}$.


\begin{equation} 
\left[\bm{\nabla_t}(\bm{\theta_t})\right]_i = s \Biggl[ \left\langle \mathcal{H}\left(\bm{\theta}_t + \bm{e}_i\frac{\pi}{4s}\right) \right\rangle \\  - \left\langle \mathcal{H}\left(\bm{\theta}_t - \bm{e}_i\frac{\pi}{4s}\right) \right\rangle \Biggr]. 
\label{eq:parameter_shift} 
\end{equation}

The iterative update to the parameter vector $\bm{\theta_{t}}$ in the $t^{th}$ iteration is given by equation \ref{eq:parameter_shift_update} where $\eta$ is the learning rate, a hyper-parameter.

\begin{equation}
    \bm{\theta_{t+1}} = \bm{\theta_{t}} - \eta \bm{\nabla_t }(\bm{\theta_{t}})
    \label{eq:parameter_shift_update}
\end{equation}

\subsubsection{Simultaneous Perturbation Stochastic Approximation gradient estimation}
\label{SPSA}

Parameter shift is a computationally expensive method for evaluating gradients because, for each parameter, two valuations of the cost function are required, and each valuation requires multiple shots of the quantum device.  By contrast, the Simultaneous Perturbation Stochastic Approximation (SPSA) technique requires only two valuations of the cost function for each iteration, independent of the number of parameters.

In SPSA \cite{spall_stochastic_1987} the iterative update to the parameter vector $\bm{\theta_{t}}$ is given by equation \ref{eq:SPSA_update} where $\bm{\nabla_t }(\bm{\theta_{t}})$ is an estimate of the gradient in the $t^{th}$ iteration.

\begin{equation}
    \bm{\theta_{t+1}} = \bm{\theta_{t}} - a_{t} \bm{\nabla_t }(\bm{\theta_{t}})
    \label{eq:SPSA_update}
\end{equation}

The $i^{th}$ component of the gradient in the $t^{th}$ iteration is given in Equation \ref{eq:SPSA_gradient} where $\mathcal{H}$ is the cost function and $\bm{\Delta _{t}}$ is a random perturbation vector with $[\bm{\Delta_t}]_i \in \{-1, 1\}$. Only two cost estimates are required to find the gradient.

\begin{equation} \left[\bm{\nabla}_t(\bm{\theta}_t)\right]_i = \frac{ \left\langle \mathcal{H}\!\left(\bm{\theta}_t + c_t \bm{\Delta}_t\right) \right\rangle - \left\langle \mathcal{H}\!\left(\bm{\theta}_t - c_t \bm{\Delta}_t\right) \right\rangle } { 2 c_t \left[\bm{\Delta}_t\right]_i } \label{eq:SPSA_gradient} \end{equation}

Both $\{a_t\}$ and $\{c_t\}$ are sequences of positive numbers converging to zero, ensuring that both the learning rate and the inputs of the gradient reduce with iteration. $\{a_t\}$ and $\{c_t\}$ are defined in equations \ref{eq:SPSA_ak} and \ref{eq:SPSA_ck}, respectively.  $A$, $c$, $\alpha$, $\eta$, and $\gamma$ are hyper-parameters and $G_0$ denotes the mean absolute value of the components of the initial gradient.  In rare cases $G_0$, as defined in equation \ref{eq:G_0}, can be evaluated at zero, so a small $\epsilon$ is added to the denominator in the calculation of the gradient to improve numerical stability.

\begin{equation}
     G_0 = \frac{1}{n} \sum_{i=1}^{n} {\left|[ \bm{\nabla}_0(\bm{\theta_{0}} \right)]_i|}
     \label{eq:G_0}
\end{equation}

\begin{equation}
    a_t = \frac{a}{t + 1 + A)^\alpha}
    \label{eq:SPSA_ak}
\end{equation}

\begin{equation}
    c_t = \frac{c}{(t+1)^{\gamma}}
    \label{eq:SPSA_ck}
\end{equation}

As coded, the SPSA algorithm above uses three average cost estimates per iteration, two to evaluate the gradient, and one to find the average cost $\langle \mathcal{H}(\bm{\theta_{t}}\rangle$ for analysis.  Inspired by Nevergrad \cite{nevergrad}, we propose Q-SPSA, which only requires one average cost estimate per iteration.  This may be beneficial with NISQ devices, where shots on quantum hardware are expensive.  In Q-SPSA, rather than using Equation \ref{eq:SPSA_gradient} to evaluate the gradient using two cost evaluations, the gradient is evaluated using current and prior iteration cost evaluations, as shown in Equation \ref{eq:Q-SPSA_gradient}.

\begin{equation} \left[\bm{\nabla}_t(\bm{\theta}_t)\right]_i = \frac{ \left\langle \mathcal{H}\!\left(\bm{\theta}_t + c'_t \bm{\Delta}_t\right) \right\rangle - \left\langle \mathcal{H}\!\left(\bm{\theta}_{t-1} + c'_{t-1}\bm{\Delta}_{t-1}\right) \right\rangle } { c'_t \left[\bm{\Delta}_t\right]_i } \label{eq:Q-SPSA_gradient} \end{equation}

and $a'_t$ and $c'_t$ are modified as shown in Equations \ref{eq:new_a} and \ref{eq:new_c} to reflect that more iterations may be necessary, as the gradient will be noisier, and hence the learning rate and the inputs of the gradient need to decay slower.

\begin{equation}
    a'_t = \frac{a}{(\lfloor \frac{t}{2} \rfloor + 1 + A)^\alpha}
    \label{eq:new_a}
\end{equation}

 \begin{equation}
    c'_t = \frac{c}{(\lfloor \frac{t}{2} \rfloor+1)^{\gamma}}
    \label{eq:new_c}
\end{equation}

Equations \ref{eq:Q-SPSA_gradient} and \ref{eq:new_a} are used to update $\bm{\theta_t}$.  

\begin{equation}
    \bm{\theta_{t+1}} = \bm{\theta_{t}} - a'_{t} \bm{\nabla_t }(\bm{\theta_{t}})
    \label{eq:SPSA_update_new}
\end{equation}

Bespoke versions of SPSA and Q-SPSA are coded in Python to allow hyper-parameter tuning.

\subsubsection{Gradient Free optimisers}
\label{grad_free}

The Nevergrad package contains many gradient free optimisers equipped with one tunable hyper-parameter $\sigma$.  The examples studied are shown in Table \ref{tab:gradient_free}

\begin{table}[H]
    \centering
    \begin{tabularx}{1.0\linewidth}{|>{\hsize=0.5\hsize\raggedright\arraybackslash}X |>{\hsize=1.5\hsize\raggedright\arraybackslash}X|}
        \hline
        \rowcolor[gray]{0.9}
        \textbf{Optimiser} & \textbf{Description}\\
         \hline
         NGOpt & Hand-crafted optimiser selection wizard provided by Nevergrad.\\
         \hline
         SPSA & Nevergrad version of SPSA where only one hyper-parameter is exposed.\\
         \hline
         OnePlusOne & Simple evolutionary algorithm with a new “offspring” selected by mutating the candidate solution and retained if successful.  If the mutation is successful the mutation strength $\sigma$ is increased.\\
         \hline
         TBPSA & Test-Based Population Size Adaptation: Generates, evaluates and compares a sample of test points, and updates best estimate.  Adjusts population size based on noisiness of results.\\
         \hline
         CMA-ES & Covariance Matrix Adaptation Evolution Strategy: Samples candidate solutions from a multivariate normal distribution and adapts the mean and covariance matrix.\\
         \hline
         PSO & Particle Swarm Optimisation:  Tracks candidate solutions (swarm of particles) and updates position and velocity based on each particle and the swarm best solution.\\
         \hline
         DE & Differential Evolution: Creates mutated vectors, randomly combines with the trial vectors (cross-over) and selects the best of the trial and mutated vector.\\
         \hline
         TwoPointsDE& Differential Evolution with a cross-over for parameters between two indices in the parameter vector.\\
         \hline
         SQP & Sequential Quadratic Programming: Replaces objective with a quadratic approximation.\\
         \hline
    \end{tabularx}
    \caption{Gradient free optimisers studied with Nevergrad}
    \label{tab:gradient_free}
\end{table}

\subsubsection{Hyper-parameter tuning}

The hyper-parameters of the optimiser described in Sections \ref{parameter_shift}, \ref{SPSA} and \ref{SGD} were varied to find the best values.  The slice described in section \ref{av} was varied for the quantum model.

\subsubsection{Warm start}
\label{warm}

A warm start may help optimisation by reducing the impact of barren plateaus (Section \ref{warm_int}).  To achieve a warm start, a greedy "nearest neighbour" classical algorithm finds a warm start cycle by iteratively selecting the nearest network location.  The corresponding warm-start bit string is found by reversing the mapping from bit string to cycle. The distance found by this greedy algorithm is compared with the results of the simulations.

The warm start is loaded onto quantum circuit 2a and 2b by initialising the $R_{XX}$ gate rotations to 0; and setting the $R_X$ gate rotations to zero, if the corresponding bit is 0, and to $\pi$ if the corresponding bit is 1; and onto the VQAi-CML model as described in Section \ref{ml_warm}  

 \subsection{Evaluation of run time dependency on network size}
\label{timing}

With Q-SPSA one average cost estimation is performed for each of the $I$ iterations.  With $n_{shot}$ shots for each iteration, the total number of shots $N_{shots}$ is given by $ I n_{shot}$.  If each shot takes time $t_{shot}$ and a time $t_{init}$ is needed before each iteration, a rough estimate of the expected algorithm completion time is $T = I n_{shot} t_{shot} + t_{init}$, which does not depend on the size of the network.  

By contrast, with parameter shift, the timing will scale less favourably with the number of locations as $\mathcal{O}(nlog_{2}n)$ where $n$ is the number of locations, because each qubit will have one or more parameters.

\section{Results}
\label{Results}

\subsection{Summary of results}

The results of the VQA and VQAi-CML models are compared with the classical Monte Carlo and Greedy methods, and the strong OR Tools classical optimisation baseline. (Section \ref{overall_results}). The VQA model finds high-quality solutions for networks of up to 17 locations and matches or outperforms Monte Carlo and Greedy heuristics, but not the OR tools strong classical server. Caching of the classical cost evaluation and SPSA reduces computational time (Section \ref{time_results}).  Warm starts (Section \ref{warm_results}) can improve the quality of the solution.  The default slicing ratio of 1 may be sub-optimal (Section \ref{slicer}).  Gray encoding may increase solution quality (Section \ref{other_parameters}).  Little impact on solution quality is found from using either the factorial and non-factorial formulation (Section \ref{other_parameters}).

\subsection{Experimental setup}

\subsubsection{Hyper-parameters used}

All simulations are, unless otherwise stated, run with the default parameters shown in Appendix \ref{parameter_settings}, found after hyper-parameter optimisation, as described in Section \ref{tunerml} and \ref{tunerqc}.

\subsubsection{Data}
 VQAi-CML and VQA evaluations are performed with networks from four to 48 locations with datasets sourced from \cite{goldsmith_beyond_2024}, \cite{reinelt_tsplibtraveling_1991}, and our own datasets manufactured by randomly sampling $n$ locations from a $100 \times 100$ grid using a uniform probability distribution.

\subsubsection{Hardware}
The simulations are run on a laptop with an Intel Core(TM) i5-10210U CPU clocked at 1.60 GHz with 8.00 GB (7.76 GB usable) RAM running Windows 11 and equipped with an NVIDIA GeForce GTX 1650 GPU.  NVIDIA CUDA is enabled for the PyTorch classical machine learning experiments.  Robustness on real quantum hardware was evaluated with the Rigetti Cepheus-1-108Q device hosted on AWS Braket, using Circuit 2b with the native device gate set described in Section \ref{qc}.     

\subsubsection{Software}
\label{software}

The notebooks and relevant Python modules are placed on GitHub \cite{noauthor_goldsmdntsp_vqc_nodate}.  For noise free runs the quantum circuits are executed in Qiskit \cite{javadi-abhari_quantum_2024} using \texttt{matrix\_product\_state} from \texttt{AerSimulator} or in AWS Braket.  For noisy runs, the circuit are transpiled onto the \texttt{FakeAuckland} digital twin of the IBM Auckland processer to use the IBM gate set, and then \texttt{qiskit\_ibm\_runtime} is used with the same fake back-end to access device-specific noise levels.  \texttt{FakeAuckland} was chosen because it has 27 qubits and a representative IBM topology.  

The Rigetti Cepheus devices used have limited connectivity, and some couplers had low fidelities. Our custom code in AWS Braket performs additional pre-run validations, ensuring that the device topology supports the specified qubit entanglement loop, and that the cumulative fidelities of all $CZ$ gates in the entanglement loop exceed a predefined threshold.  Classic machine learning circuits are coded in PyTorch \cite{paszke_pytorch_2019}. 

Google OR-Tools provided a strong classical optimisation baseline and was configured using the \texttt{PATH\_CHEAPEST\_ARC} first-solution strategy to generate an initial tour, followed by \texttt{GUIDED\_LOCAL\_SEARCH} to iteratively improve the solution. The search was allowed to proceed for a fixed time limit (60 seconds) before returning the best tour found.

\subsubsection{Monte Carlo benchmark}
\label{montecarlo}

The simulation results of both VQA and VQAi-CML models are benchmarked using a Monte Carlo evaluation.  A Python method coded in the caching class (Section \ref{cache}) counts the number of bit strings sampled by our models and the same number of randomly generated bit strings are evaluated to find the minimum distance. The number of bit strings sampled varies, because each shot might return from 1 to $2^q$, bit strings, where $q$ is the number of qubits.

\subsubsection{Solution quality metric}

The quality of the solution is defined as $Q_{sol}  =  \frac{D_{best}}{D_{sim}}$. The minimum simulation distance found ($D_{sim}$) is compared with the best known minimum distance ($D_{best}$) taken from the datasets \cite{goldsmith_beyond_2024} and \cite{reinelt_tsplibtraveling_1991}, or by brute force enumeration over all cycles for our own datasets.  A solution quality of $Q_{sol} = 1$ represents a solution where the shortest distance found is optimal, and a value less than 1 represents a suboptimal solution. Since the quality of the solution is often close to 1, it is sometimes more helpful to consider the solution error $E_{sol}  =  1 - Q_{sol}$.

\subsubsection{Estimated sample error}
The runs are stochastic because of the inherent probabilistic nature of quantum computing and stochastic gradient descent, where used. Here, $r$ runs are carried out for the same set of parameters, an estimate of the sample error of the mean 
$$SEM ={\frac {\sigma }{\sqrt {r-1}}}$$ 
is provided where $\sigma$ denotes the standard deviation of the observed sample. The error bars in the figures correspond to $\pm SEM$, calculated for the independent optimisation runs performed for each network instance and model. 

\subsection{Model comparison}
\label{overall_results}

\begin{figure}[htbp]
    \centering
    \includegraphics[width=1.0\linewidth]{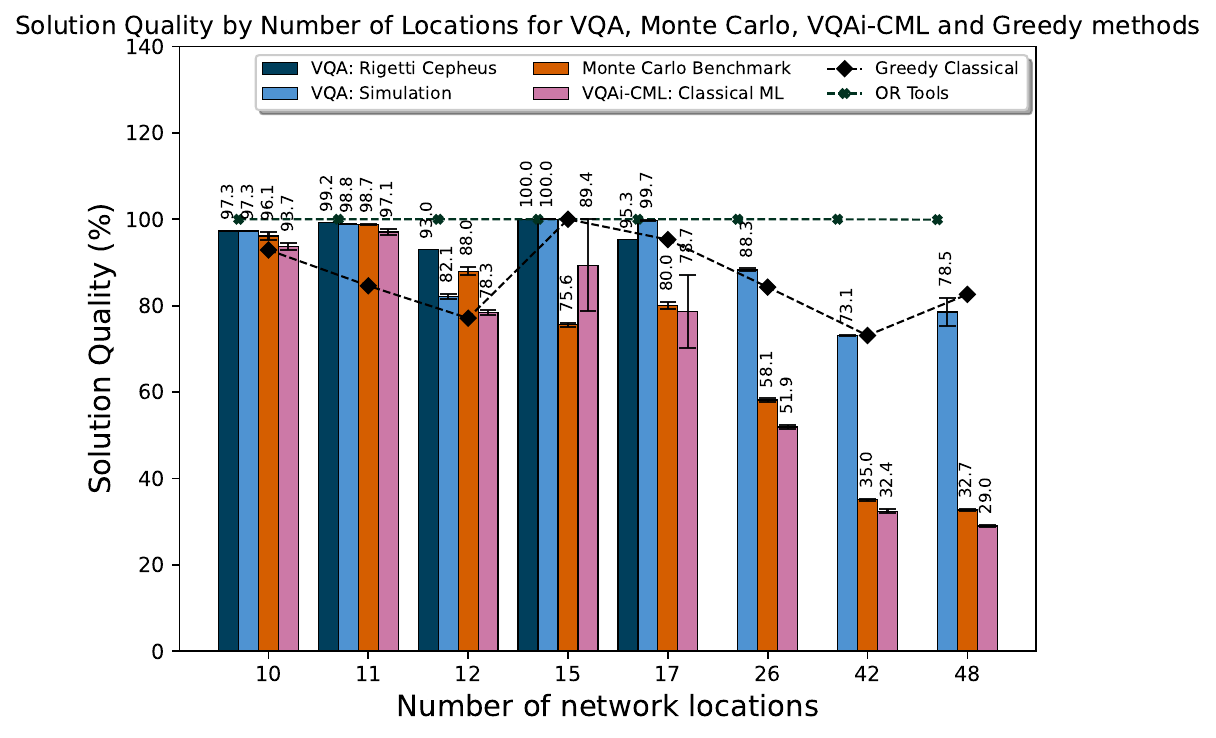}
    \caption{Comparison between the average solution quality for networks with different number of locations for VQA hardware and simulations, a Monte Carlo baseline, the VQAi-CML model, a greedy classical heuristic and Google OR Tools}
    \label{fig:result_summary}
\end{figure}

The quality of the solution by network size is shown in Figure \ref{fig:result_summary} and Table \ref{tab:overall_summary}.  The runs shown are with a warm start.  The VQA runs use Circuit 2b, and the CMA-ES optimiser, ensuring comparability.

\begin{table}[H] 
\centering 
\begin{threeparttable}
    \begin{tabularx}{1.0\linewidth}{|C|C|C|C|C|C|C|}
        \hline
         \textbf{Locations} &  
         \textbf{VQA: Rigetti} \footnotemark[1] & 
         \textbf{VQA: Sim'n} \footnotemark[2] &
         \textbf{Monte Carlo} \footnotemark[3]& 
         \textbf{VQAi-CML:} \footnotemark[4]&
         \textbf{Greedy} \footnotemark[5]&
         \textbf{OR Tools} \footnotemark[6]\\ 
         \hline
    10 & $97.3$  & $97.3\pm0.0$&$ 96.1\pm0.9$&$93.7\pm0.8$ &$ 92.9$&$100.0$\\ \hline
    11 & $99.2$  & $98.8\pm0.0$&$ 98.7\pm0.1$&$97.1\pm0.8$ &$ 84.6$&$100.0$\\ \hline
    12 & $93.0$  & $82.1\pm0.6$&$ 88.0\pm0.9$&$78.3\pm0.6$ &$ 77.1$&$100.0$\\ \hline
    15 & $100.0$ &$100.0\pm0.0$&$ 75.6\pm0.6$&$89.4\pm10.6$&$100.0$&$100.0$\\ \hline
    17 & $95.3$  & $99.7\pm0.1$&$ 80.0\pm0.5$&$78.7\pm8.4$ &$ 95.3$&$100.0$\\ \hline
    26 & n/a     & $88.3\pm0.3$&$ 58.1\pm0.4$&$51.9\pm0.5$ &$ 84.3$&$100.0$\\ \hline
    42 & n/a     & $73.1\pm0.0$&$ 35.0\pm0.2$&$32.4\pm0.5$ &$ 73.1$&$100.0$\\ \hline
    48 & n/a     & $24.9\pm1.0$&$ 10.4\pm0.1$&$ 9.2\pm0.1$ &$ 26.2$&$ 99.9$\\ \hline
    \end{tabularx}
    \begin{tablenotes} 
    \footnotesize 
    \item[1] VQA Rigetti Cepheus-1-108Q: 1,250 iterations are used apart from runs with 10 and 15 locations, which only needed 700 and 5 iterations, respectively. Only one run is carried out, so no errors are estimated. 
    \item [2]{VQA simulations: run with 1,250 iterations}
    \item[3]{Monte Carlo benchmark: Section \ref{montecarlo}}
    \item[4]{VQAi-CML: Section \ref{ml}}
    \item[5]{Greedy Classical Heuristic: Section \ref{warm}}
    \item[6]{Google OR-Tools: Section \ref{software}}
    \end{tablenotes}
     \caption{Comparison between the average solution quality for networks with different number of locations for VQA hardware and simulations, a Monte Carlo baseline, a VQAi-CML model, a greedy classical heuristics and Google OR Tools.  The error shown is the SEM for the independent optimisation runs performed for each network instance and model.}

    \label{tab:overall_summary}
\end{threeparttable} 
\end{table}

\FloatBarrier

In simulations of up to 9 qubits, including noisy simulations using the \texttt{FakeAuckland} IBM fake backend, optimal solutions are found.   Noisy simulations of larger circuits are omitted from the evaluation due to their prohibitive execution times. Furthermore, the availability of results from real quantum hardware renders such simulations unnecessary.  For networks of up to 26 locations, the VQA matches or outperforms the classical Monte Carlo, Greedy, and VQAi-CML methods, with the sole exception of the 12-location simulation, for which the solution quality is marginally below the Monte Carlo benchmark.  For problem instances up to 26 locations (94 qubits), stable convergence is observed across independent runs, but the convergence speed, and hence the quality of the solution, declines for larger networks with 42 and 46 locations (183 and 219 qubits, respectively).  

The Rigetti hardware run outperforms the classical methods for 12 locations, and for 17 locations has similar solution quality to the greedy heuristic, and a lower solution quality than in simulation.  The cumulative $CZ$ gate fidelity around the 50 qubit-loop needed for 17 locations is only just above 50\%, and noise may have made training difficult for larger network sizes.  

For networks with 15 and 17 locations, the Greedy heuristic produces high-quality solutions. Notably, for the 17-location instances, the VQA is able to improve on the Greedy heuristic in simulation. In contrast, the VQAi-CML model generally finds lower quality solutions and may require further investigation and refinement.  OR-Tools finds optimal solutions in 60 seconds for networks containing up to 42 locations and achieved a near-optimal solution quality of 99.9\% for the 48-location instance.

\subsection{Component analysis}

The model components, including optimisers, warm and cold start, slicing ratios, circuits, Gray encoding, and formulation, are systematically varied to find their impact on model performance.

\subsubsection{Optimisers}
\label{optimiser_results}

\begin{figure}[htbp]
    \centering
    \includegraphics[width=1\linewidth]{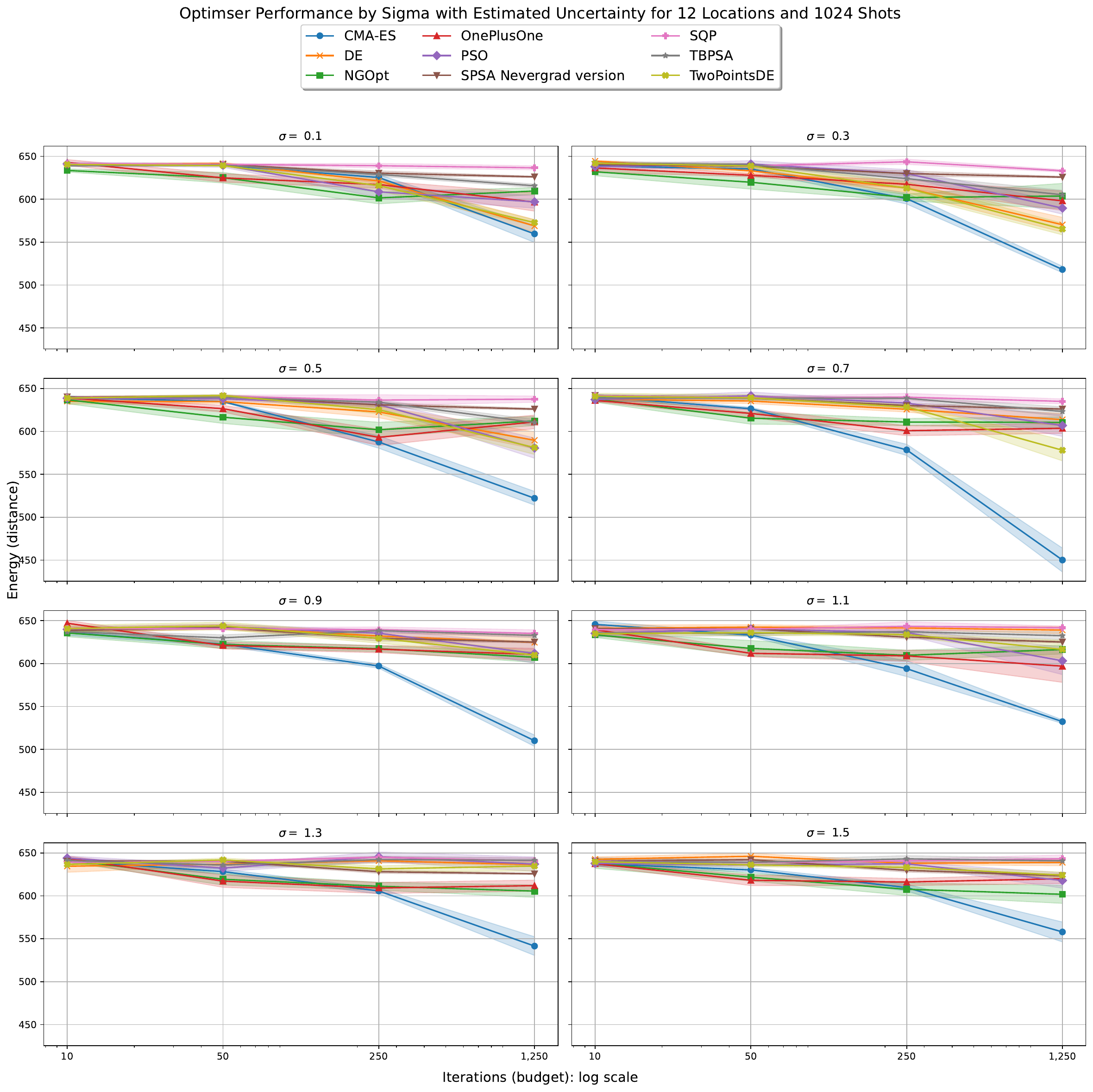}
    \caption{Optimiser performance for different Nevergrad optimisers showing that the CMA-ES converges better to low cost solutions.}
    \label{fig:optimisers}
\end{figure}

Figure \ref{fig:optimisers} compares the nine different Nevergrad optimisers discussed in Section \ref{grad_free} and assesses the impact of different values of $\sigma$, a tuneable hyper-parameter .  It can be seen that the optimiser that converges best to lower cost solutions, as evidenced by the lowest cost after 1,250 iterations, is CMA-ES with $\sigma$ = 0.7.  

\begin{figure}[htbp]
    \centering
    \includegraphics[width=1\linewidth]{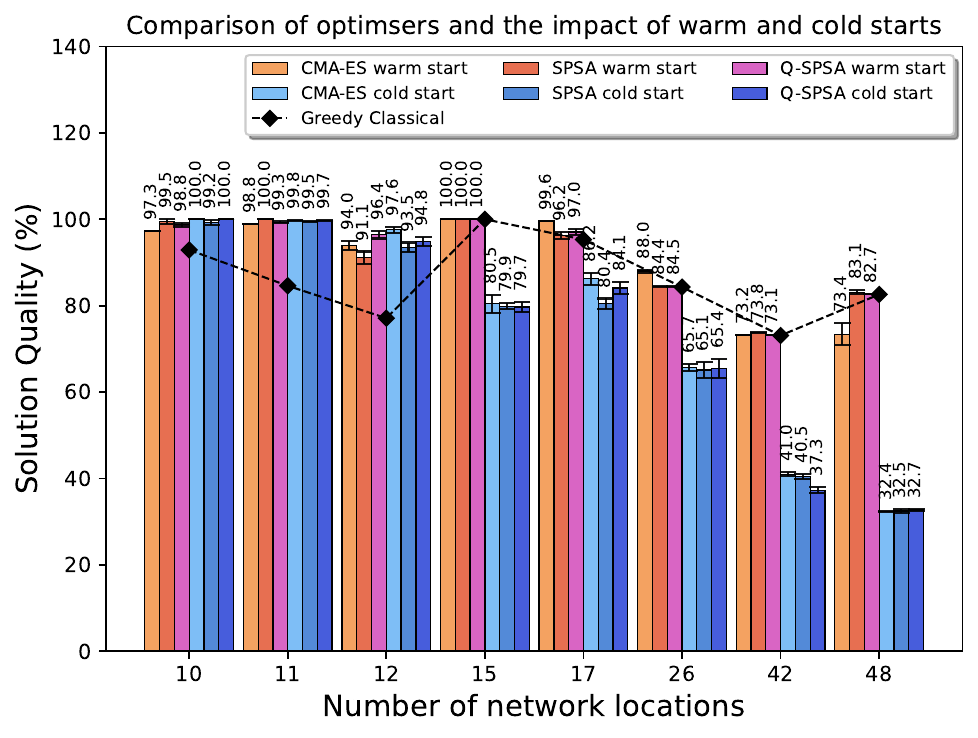}
    \caption{Comparison of CMA-ES, SPSA and Q-SPSA for warm and cold start}
    \label{fig:optimisers_comp}
\end{figure}

The Nevergrad package only exposes one SPSA hyper-parameter $\sigma$, whereas SPSA and Q-SPSA (section \ref{SPSA}) have five hyper-parameters that can be tuned: $A, c, \alpha, \eta$ and $\gamma$.  To enable hyper-parameter tuning, bespoke versions of SPSA and Q-SPSA are coded.  Since SPSA uses three evaluations of the average distance per iteration, and Q-SPSA and CMA-ES use one, SPSA is run with 417 iterations, whereas Q-SPSA and CMA-ES are run with 1,250 iterations to give a fair comparison.  Figure \ref{fig:optimisers_comp} shows the performance of the three different optimisers for warm and cold starts using circuit 2a.

It can be seen that although the three optimisers have similar performance, CMA-ES performs slightly better in hot starts with larger networks ($99.6 \% \pm0.0$ for CMA-ES versus $97.0 \% \pm0.7$ for Q-SPSA for 17 locations), with a Welch $t$-test indicating that the improvement was statistically significant ($p = 0.005$).

\subsubsection{Warm start}
\label{warm_results}

Figure \ref{fig:optimisers_comp} shows that the impact of a warm start depends on how close the greedy warm start initialisation is to the optimal solution.  Comparison of the Hamming distance, defined as the number of differing bit positions between two binary strings, suggests that the warm-start initialisation may not always lie close to the optimal solution for certain network instances. Table \ref{tab:hamming} in Appendix \ref{appendix_results} shows the Hamming distance between the warm start initialisation binary string and the optimum binary string, found by brute-force.  No results are provided for larger networks than 15 locations as the computational cost of brute force was prohibitive.

\subsubsection{Slicing}
\label{slicer}

\begin{figure}[H]
    \centering
    \includegraphics[width=1.0\linewidth]{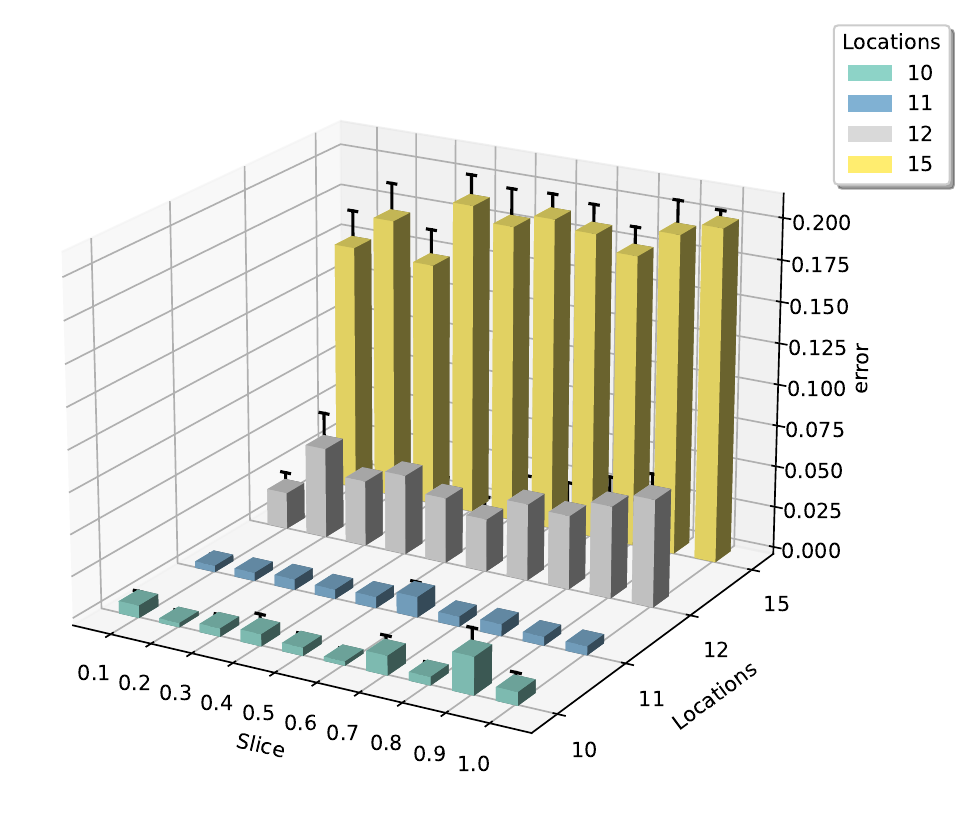}
\caption{Average VQA error by slicing for different network sizes.}
    \label{fig:slice}
\end{figure}

Figure \ref{fig:slice} illustrates the average error by location for different slice ratios (Section \ref{av}).  The results indicate that smaller slice ratios than 1.0 can sometimes produce lower errors. The Welch \(t\)-test indicated that the improvement is not statistically significant, for example, errors of $17.6 \% \pm1.3$ are found for a slice of 0.8 versus $20.3 \% \pm0.7$ for a slice of 1.0 with 15 locations ($p = 0.12$).  At least ten data runs are performed for each slice and location combination.

\subsubsection{Other components evaluated}
\label{other_parameters}

 A summary of results for other components is shown in Table \ref{tab:component}. The baseline uses 15 locations, Circuit 2a, a cold start, the Q-SPSA optimiser, no slicing, non-factorial formulation and no Gray encoding,

\begin{table}[H]
    \centering
    \begin{tabularx}{0.8\linewidth}{|C|C|C|}
        \hline
         \rowcolor[gray]{0.9}
         Configuration& Impact assessed  & Solution Quality (\%)\\
         \hline
         \textbf{Baseline} & n/a  & \bm{$78.7\pm0.6$} \\
         \hline
         Circuit 1 & Circuits & $79.5\pm0.9$\\
         \hline
         Circuit 2a & Circuits & $78.7\pm0.6$ \\
         \hline
         Circuit 2b & Circuits & $79.1\pm1.1$\\
         \hline
         Circuit 3 & Circuits & $79.0\pm1.2$\\
         \hline
         Circuit 4 & Circuits & $79.6\pm2.0$\\
         \hline
         Circuit 5 & Circuits & $79.1\pm1.1$\\
         \hline
         Gray = True & Gray encoding & $83.4\pm1.2$\\
         \hline
        Factorial formulation & Formulation & $78.3\pm0.7$\\
         \hline
    \end{tabularx}
    \caption{Summary of component analysis}
    \label{tab:component}
\end{table}

The performance of the quantum circuits is broadly comparable, as shown in Table \ref{tab:component}. Circuits 2a and 2b are selected for further study, as they support both warm-start initialisation and entanglement.

Gray encoding produces a higher average solution quality than baseline ($83.4 \%\pm1.2$ versus $78.7 \% \pm0.7$). The observed difference suggests that Gray encoding may provide a performance benefit over baseline, with a Welch $t$ -test indicated that the improvement was statistically significant ($p = 0.0017$).

 It can be seen that there is no statistically significant difference in the solution quality of the factorial and non-factorial formulations ($78.3 \%\pm0.7$ versus $78.7 \% \pm0.6$).  In addition to solution quality, the number of qubits required by each formulation is important.  Table \ref{tab:form1} in Appendix \ref{appendix_results} shows the qubits required for different network sizes for the factorial and non-factorial formulation.  It can be seen that for small location sizes the non-factorial formulation requires fewer qubits, and this is reversed for networks with more than 15 locations.

\subsubsection{Computational time decreases}
\label{time_results}

The use of caching and gradient free optimisers, such as SPSA, significantly reduce overall run times, for example, from more than 9 minutes to seven seconds, as shown in Table \ref{tab:time}.  The table shows the average timings for four runs with eight locations with SPSA and Parameter Shift gradient descent, as described in Section \ref{optimisation}.   Caching the classical cost evaluation of the distance for bit string significantly reduces run-time, for example from 37 to 7 seconds with SPSA and was used by default.

\begin{table}[H]
\centering
\begin{tabularx}{0.6\linewidth}{|C|C|C|}
 \hline
 \rowcolor[gray]{0.9}
 & \textbf{SPSA} & \textbf{Parameter shift}\\
 \hline
 No caching  & 37 secs  & 585 secs \\
  \hline
 Caching  & 7 secs & 217 secs \\
\hline
\end{tabularx}
\caption{Table  showing caching significantly reduces run times, for different gradient descent methods, for a network with eight locations.  SPSA is significantly quicker than parameter shift.}
\label{tab:time}
\end{table}

\subsubsection{Hyper-parameter tuning results for VQAs}
\label{tunerqc}

For the parameter-shift optimiser, the learning rate $\eta$ is varied whilst keeping $s$ constant.  Then $s$ is varied using the best value found for $\eta$.  The standard choice of 0.5 appears to be satisfactory.  For the SPSA and Q-SPSA optimiser, $A$ and $\eta$ are varied in a logarithmic grid search.  Once the best values for these were found, $\alpha$, $\gamma$ and $c$ are also varied.  The optimum hyper-parameters found are given in Appendix \ref{parameter_settings}.  High quality solutions with standard values of $\alpha$, $\gamma$, and $c$ were found.

The parameter shift optimiser is found to take a long time to run (see Table \ref{tab:time}), as expected because multiple average cost evaluations are needed, one for each parameter.  Moreover, it is often noted that the cost function did not converge to a set value.  Because gradient-free methods such as Q-SPSA gives similar high quality solutions and are much faster than parameter shift, Q-SPSA is used as the default parameter unless stated otherwise.

\subsubsection{Hyper-parameter tuning results for VQAi-CML}
\label{tunerml}
For VQAi-CML, SGD and Adam are investigated, as described by  \cite{Goodfellow-et-al-2016} in the context of deep learning.  For SGD, changes in momenta $\beta$, learning rate $\eta$, and weight decay $\lambda$ are investigated.  For Adam a standard momentum $\beta_1$ of 0.9 is used and a logarithmic grid search is performed for the learning rate $\eta$ and the weight decay $\lambda$.  In both cases it was important to use a non-zero weight decay, this may be because it is important to keep the weights small for stability in training, because the $Sine$ activation layer maps multiple of $2\pi$ to the same value. The best value for $\sigma_v$, the variability of the initial variables in the warm start for the VQAi-CML model is found to be 0.05.

\subsubsection{VQAi-CML model}

Section \ref{MLr} describes how adding more layers to the VQAi-CML model does not significantly affect the quality of the solution,  and changing the number of binary vectors, or shots, in each epoch by the VQAi-CML model improves the quality of the solution, at the cost of longer run times.

\section{Conclusions and future work}
\label{Conclusion}

The proposed VQA model achieves high-quality solutions in both simulation networks containing up to 48 locations and in real quantum hardware for networks containing up to 17 locations. To the best of our knowledge, existing studies of the circuit model paradigm have been limited to simulations involving at most eight locations, with the reported solutions remaining sub-optimal \cite{bai_quantum_2025, bourreau_indirect_2023}. These results demonstrate the scalability of the proposed approach relative to previously reported work.

The VQA architecture uses shallow circuits and a relatively small number of qubits, reducing the dimensionality of the optimisation landscape and consequently mitigating barren plateau effects while enhancing noise resilience.
As such, the conditions under which barren plateaus have been rigorously established may not apply.  For problem instances of up to 26 locations (94 qubits), we observe stable convergence across independent runs, with no empirical indication of vanishing gradients, suggesting that landscape trainability is retained at these scales.

Although this work is motivated by near-term quantum hardware constraints, its relevance is not limited to the NISQ era. In particular, the penalty‑free encoding of problem constraints addresses a fundamental algorithmic challenge that persists irrespective of hardware maturity. Even in fault‑tolerant regimes, unnecessary penalty terms increase circuit and optimisation landscape complexity, and a warm start may increase solution quality.  Although the variational optimisation techniques employed here may be enhanced in fault-tolerant architectures, the core encoding framework developed in this work provides a foundation for future advances.

In addition, a quantum-inspired classical machine learning model (VQAi-CML) is implemented as a benchmarking baseline. Although this model does not exceed the performance of the VQA, it provides a useful point of comparison for evaluating the advantages and limitations of the quantum approach. Further points of comparison are provided by the Monte Carlo and Greedy heuristics. Across the evaluated instances, the VQA model almost always achieves solution quality that is comparable to, or better, than these benchmarks, indicating an ability to effectively explore the solution space. Nevertheless, Google OR-Tools consistently achieves higher-quality solutions and therefore provides a strong classical benchmark.

Our findings suggest several avenues for future research. Warm-start initialisation based on a greedy nearest-neighbour heuristic improves performance when the initial solution was sufficiently close to the optimum. Given the availability of more sophisticated classical approximation algorithms for the TSP \cite{papadimitriou_combinatorial_1998} future work could investigate alternative warm-start strategies and assess their effectiveness in mitigating barren plateau effects \cite{larocca_barren_2025, cerezo_does_2024}. Further improvements may also be possible for VQAi-CML through the use of alternative weight-initialisation schemes. In particular, the sinusoidal initialisation proposed by \cite{sitzmann_implicit_2020} for sine activations may offer advantages over the Kaiming-He uniform initialisation \cite{he_delving_2015} employed by default in PyTorch linear layers. Finally, Markovian bit-string generators are capable of producing correlated samples \cite{shamshad_first_2005}; consequently, future studies could investigate Markov chain-based generators to determine whether they can improve upon the performance achieved by the model studied.

\FloatBarrier

\section{Acknowledgments}

This project was funded and supported by the UK National Quantum Computer Centre [NQCC200921], which is a UKRI Centre and part of the UK National Quantum Technologies Programme (NQTP). Many thanks to Manuel Schnaus for helpful input on his formulation of TSP and to Alexander Makarovskiy for valuable input on Nevergrad.

IBM\textsuperscript{\textregistered} and Qiskit\textsuperscript{\textregistered} are registered trademarks of International Business Machines Corp.  NVIDIA\textsuperscript{\textregistered}, GeForce\textsuperscript{\textregistered} and CUDA\textsuperscript{\textregistered}  are registered trademarks of NVIDIA. Intel\textsuperscript{\textregistered}  is a registered trademark of Intel Corporation.  Rigetti\textsuperscript{\textregistered} and Cepheus\textsuperscript{\textregistered} are registered trademarks of Rigetti.  Google\textsuperscript{\textregistered}is a registered trademark of Google LLC.  AWS Braket\textsuperscript{\textregistered} is a registered trademark/service mark of Amazon

\section{References}

\bibliographystyle{unsrt} 
\bibliography{references}

\appendix
\label{appendix}
\section{Appendix - Parameter settings}
\label{parameter_settings}

\begin{table}[H]
    \centering
    \begin{tabularx}{1\textwidth}{|L|C|L|}
        \hline
        \rowcolor[gray]{0.9}
        \textbf{Parameter} &\textbf{Default} & \textbf{Notes} \\ \hline
        Formulation & Original &  Section \ref{other_parameters} \\ \hline
        Warm start   & False & Section \ref{warm_results}\\ \hline
        Gray  & False & Section  \ref{other_parameters} \\  \hline
        Slice & 1.0   & Section \ref{slicer} \\ \hline
    \end{tabularx}
    \caption{General default parameters used}
    \label{tab:General_defauls}
\end{table}

\begin{table}[H]
    \centering
    \begin{tabularx}{1\textwidth}{|L|C|L|}
        \hline
        \rowcolor[gray]{0.9}
        \textbf{Parameter} &\textbf{Default} & \textbf{Notes}\\ 
        \hline
        Circuit    & 2a &  Section \ref{other_parameters}\\ 
        \hline
        Shots  &1024 & \\ 
        \hline
        Gradient type & Q-SPSA &Section \ref{tunerqc}\\ 
        \hline
        Q-SPSA: A & 100 & Section \ref{tunerqc}\\ 
        \hline
        SPSA: A & 25 & Section \ref{tunerqc}\\ 
        \hline
        Q-SPSA: Learning rate $(\eta)$ & 0.005 & Section \ref{tunerqc}\\ 
        \hline
        SPSA: Learning rate $(\eta)$ & 0.1 & Section \ref{tunerqc}\\ 
        \hline
        SPSA and Q-SPSA: c & $\pi/10 =  0.314$ & Section \ref{tunerqc}\\ 
        \hline
        SPSA and Q-SPSA: $\alpha$ & 0.602 & Section \ref{tunerqc}\\ 
        \hline
        SPSA and Q-SPSA: $\gamma$ & 0.101 &  Section \ref{tunerqc}\\ 
        \hline
        CMA-ES: $\sigma$ & 0.7 &  Section \ref{tunerqc}\\ 
        \hline
        Parameter shift: Learning rate $(\eta)$  & 0.1 & Section \ref{tunerqc}\\  
        \hline
        Parameter shift: s  &0.5 & Section \ref{tunerqc}\\ 
        \hline   
    \end{tabularx}
    \caption{VQA default parameters used}
    \label{tab:VQA_defaults}
\end{table}

\begin{table}[H]
    \centering
    \begin{tabularx}{1\textwidth}{|L|C|L|}
        \hline
        \rowcolor[gray]{0.9}
        \textbf{Parameter} &\textbf{Default} & \textbf{Notes} \\
        \hline
        ML input vector no warm start& Zeros &  Section \ref{MLr}\\
        \hline
        Number of input vectors (shots) & 64 & Section \ref{MLr}\\
        \hline
        Gradient type & SGD & Section \ref{tunerml}\\
        \hline
        $\sigma$ & 0.05 &  Section \ref{tunerml} \\
        \hline
        SGD: LR ($\eta$) & $2 \times 10^{-5}$ & Section \ref{tunerml}\\
        \hline
        SGD: Momentum $\beta$& 0.8 & Section \ref{tunerml}\\
        \hline
        SGD Weight decay $\lambda$ & 0.0006 & Section \ref{tunerml}\\ 
        \hline
        Adam: LR ($\eta$) & 0.001 & Section \ref{tunerml}\\ 
        \hline
        Adam: Momentum $\beta_1$  & 0.9 & Section \ref{tunerml}\\
        \hline
        Adam Weight decay $\lambda$ & 0.0032 & Section \ref{tunerml} \\
        \hline       
    \end{tabularx}
    \caption{VQAi-CML Default parameters used}
    \label{tab:ML_defaults}
\end{table}

\section{Appendix - Detailed results}
\label{appendix_results}

\FloatBarrier

\subsection{VQA results}

\begin{table}[H]
    \centering
    \begin{tabularx}{0.6\linewidth}{|C|C|C|}
    \hline
    \rowcolor[gray]{0.9}
    \textbf{Locations} & \textbf{Non-factorial Qubits} & \textbf{Factorial Qubits}  \\ \hline
    5  & 5   & 7   \\ \hline
    10 & 21  & 22  \\ \hline
    15 & 41  & 41  \\ \hline
    20 & 64  & 62  \\ \hline
    25 & 89  & 84  \\ \hline
    30 & 114 & 108 \\ \hline
    35 & 141 & 133 \\ \hline
    40 & 171 & 160 \\ \hline
    45 & 201 & 187 \\ \hline
    50 & 231 & 215 \\ \hline
    55 & 261 & 243 \\ \hline
    60 & 291 & 273 \\ \hline
    \end{tabularx}
    \caption{Qubits required for different location sizes for different network sizes for the factorial and non-factorial formulation.}
    \label{tab:form1}
\end{table}

 \begin{table}[H]
    \centering
    \begin{tabularx}{\linewidth}{|C|C|C|C|C|C|}
        \hline
        \rowcolor[gray]{1.0}
        \textbf{Locations} & \textbf{Qubits} & \textbf{Best dist.} & \textbf{Warm start dist.} & \textbf{Warm start sol. qual.} & \textbf{Lowest hamming dist.} \\
        \hline
        4         & 3      & 21.0      & 21.0           & 100.0   & 0 \\ \hline
        5         & 5      & 19.0      & 21.0           & 90.5   & 2 \\ \hline
        6         & 8      & 241.0     & 279.6          & 86.2   & 5 \\ \hline
        7         & 11     & 276.2     & 314.8          & 87.7   & 5 \\ \hline
        8         & 14     & 277.2     & 315.8          & 87.8   & 8 \\ \hline
        9         & 17     & 286.7     & 339.8          & 84.4   & 9 \\ \hline
        10        & 21     & 290.2     & 312.3          & 92.9   & 8 \\ \hline
        11        & 25     & 253.0     & 299.0          & 84.6   & 9  \\ \hline
        12        & 29     & 297.2     & 385.5          & 77.1   & 10  \\ \hline
        15        & 41     & 291.0     & 291.0          & 100.0   & 0  \\ \hline
    \end{tabularx}
    \caption{Table showing that the Hamming distance between the warm start and optimum cycle can be significant.  The table shows distances, warm start solution quality in percent, and Hamming distances for various locations and qubits, showing that the warm start initialisation is not always close to the optimal string.}
    \label{tab:hamming}
\end{table}

\FloatBarrier

\subsection{VQA-inspired Classical machine learning model (VQAi-CML)}
\label{MLr}

The average error for the VQAi-CML, with optimal parameters, by location and number of layers is plotted in Figure \ref{fig:ML1} in Appendix \ref{appendix_figures} and the corresponding solution quality is provided in Table \ref{tab:ML3}.  The results show that the number of layers has little impact on the final quality of the solution.

\begin{table}[H]
    \centering
    \begin{threeparttable}
    \begin{tabular}{|c|c|c|c|c|}
        \hline
        \rowcolor[gray]{0.9}
        \textbf{Locations} & \textbf{1 layer} & \textbf{2 layers} & \textbf{3 layers} & \textbf{4 layers}\\ \hline
         4 & $100.0$    & $100.0 $ & $100.0 $ & $100.0  $\\ \hline
         5 & $100.0 $   & $100.0 $ & $100.0 $ & $100.0  $\\ \hline
         6 & $94.0 $    & $94.0 $  & $100.0 $ & $99.5  $\\ \hline
         7 & $89.7 $    & $89.0 $  & $89.7 $  & $88.2  $\\ \hline
         8 & $99.8 $    & $83.3 $  & $97.1 $  & $89.7  $\\ \hline
         9 & $83.3 $    & $92.1 $  & $81.5 $  & $82.1  $\\ \hline
         10 & $81.1 $   & $77.1 $  & $74.1 $  & $72.0  $\\ \hline
         11 & $89.4 $   & $87.5 $  & $93.4 $  & $88.5  $\\ \hline
         12 & $64.8 $   & $58.8 $  & $63.5 $  & $68.1  $\\ \hline
         15 & $66.1 $   & $69.0 $  & $68.8 $  & $65.6  $\\ \hline
         17 & $76.6 $   & $70.0 $  & $71.7 $  & $68.9  $\\ \hline
         26 & $94.1 $   & $94.1 $  & $94.1 $  & $94.1  $\\ \hline
         42 \footnotemark[1] & $100.0 $  & $100.0 $ & $100.0 $ & $100.0 $\\ \hline
         48 & $9.3 $    & $9.2 $   & $8.9 $   & $9.4   $\\ \hline
         \hline
         \textbf{Overall} & \bm{$82.0 $}    & \bm{$80.3 $}   
         & \bm{$81.6 $}   & \bm{$80.4   $}\\ \hline
    \end{tabular}
    \begin{tablenotes} 
    \footnotesize 
    \item[1] {Optimum results are found for the VQAi-CML model for the network with 42 locations with a cold start.  Plotting the relevant data \texttt{set dantzig42\_xy.csv} showed that ordering the locations in numerical order gives the optimum solution: this is an easy solution for our formulation to find.}
    \end{tablenotes}
    \caption{There is little difference between the solution quality, in percent, for the VQAi-CML models, with 1, 2, 3 and 4 layers with networks of different number of locations.  No sample error is estimated as most combinations were only run once.}
    \label{tab:ML3}
    \end{threeparttable}
\end{table}

\begin{figure}[h]
    \centering
    \includegraphics[width=1.0\linewidth]{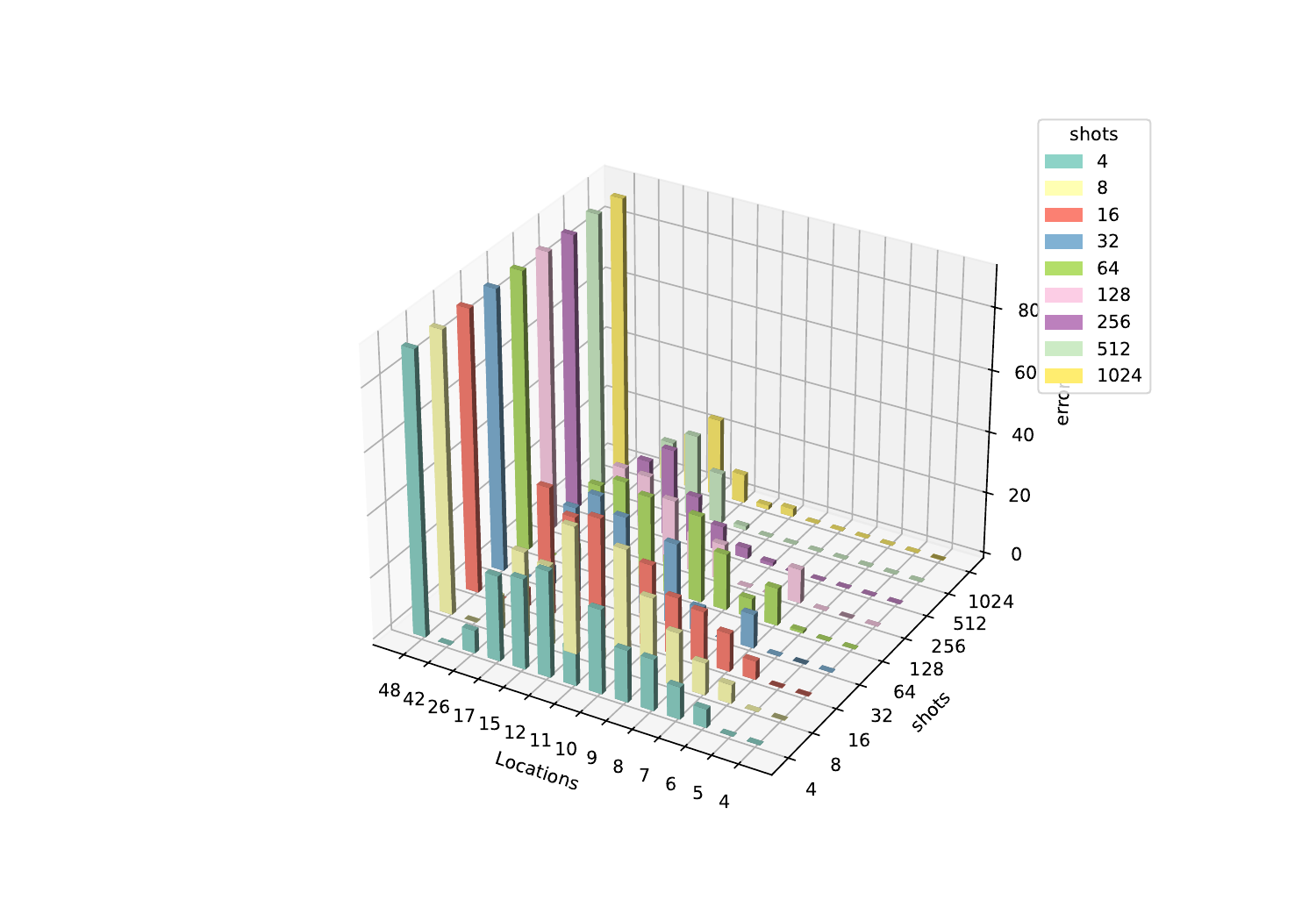}
    \caption{Solution error in \%age for VQAi-CML with the number of binary input vectors (shots, or mini-batches) against the number of locations, showing that the error reduce as the numbers of the binary input vectors increase}
    \label{fig:shots}
\end{figure}

The number of input vectors were varied for each epoch for different numbers of locations.  The summary results are plotted in Figure \ref{fig:shots}, shown in summary in Table \ref{tab:minibatch}  with detailed results in Table \ref{tab:bin_input}.  Although the quality of the solution increases with the number of input vectors, so does the run time, which varies approximately linearly with the number of input vectors.  As discussed above, using more shots improves the fraction of the solution space investigated, and this might explain the increase in solution quality.  64 input vectors per epoch were used for hyper-parameter optimisation as a good compromise between solution quality and execution time.  The results quoted in the general result summary in Section \ref{overall_results} are for 1,024 input vectors.  The warm start results in Section \ref{warm_results} in Appendix \ref{appendix_figures} and the impact of the VQAi-CML layers above are based on runs with 64 input vectors. 

\begin{table}[H]
    \centering
    \begin{tabularx}{\linewidth}{|C|C|C|C|}
        \hline
        \rowcolor[gray]{0.9}
        \textbf{No. of input vectors} & \textbf{Solution found on iteration (av.)} & \textbf{Av solution quality} & \textbf{Av. Execution time (secs)}\\
        \hline
          4  & 79.3 & 80.1 & 42 \\  \hline
          8  & 49.1 & 78.0 & 84 \\  \hline
         16  & 41.2 & 78.6 & 166 \\ \hline
         32  & 32.2 & 82.3 & 331 \\ \hline
         64  & 37.9 & 81.2 & 658 \\ \hline
        128  & 41.2 & 85.0 & 1,346 \\\hline
        256  & 49.4 & 87.2 & 2,854 \\ \hline
        512  & 58.8 & 88.1 & 8,090 \\ \hline
        1,024& 72.3 & 92.3 & 10,082 \\ \hline
    \end{tabularx}
    \caption{The overall average solution quality in percent and elapsed time increases as the number of input binary vectors (mini-batches) increases for VQAi-CML, at the expense of enhanced execution time.}
    \label{tab:minibatch}
\end{table}

The input vectors were initialised to values of 0, or 0.5.  Detailed results are shown for 64 and 256 input vectors, respectively, in Tables \ref{tab:init64} and \ref{tab:init256} showing that although the solution quality is slightly better for initialisations with 0.5 with 64 input vectors, this result does generalise to 256 input vectors.  

\begin{table}[H]
    \centering
    \begin{tabularx}{0.6\linewidth}{|C|C|C|}
        \hline
        \rowcolor[gray]{0.9}
        \textbf{Locations} & \textbf{Input vector with zeros} & \textbf{Input vector with 0.5} \\ \hline
          4  & $100.0 \pm 0.0$ & $100.0 \pm 0.0$ \\ \hline
          5  & $100.0 \pm 0.0$ & $100.0 \pm 0.0$ \\ \hline
          6  & $99.3  \pm 0.7$ & $96.3  \pm 2.5$ \\ \hline
          7  & $87.9  \pm 1.2$ & $93.1  \pm 2.8$ \\ \hline
          8  & $93.9  \pm 3.8$ & $94.5  \pm 3.5$ \\ \hline
          9  & $81.8  \pm 2.8$ & $91.5  \pm 2.9$ \\ \hline
          10 & $71.9  \pm 2.7$ & $83.0  \pm 3.0$ \\ \hline
          11 & $88.2  \pm 0.4$ & $92.8  \pm 0.3$ \\ \hline
          12 & $70.4  \pm 4.5$ & $80.9  \pm 0.7$ \\ \hline
          15 & $67.9  \pm 1.7$ & $70.0  \pm 0.8$ \\ \hline
          17 & $71.5  \pm 0.6$ & $72.1  \pm 1.8$ \\ \hline
          26 & $94.1  \pm 0.9$ & $92.5  \pm 0.7$ \\ \hline
          42 & $100.0 \pm 0.0$ & $100.0 \pm 0.0$ \\ \hline
          48 & $9.4   \pm 0.1$ & $9.2   \pm 0.0$  \\ \hline
          \hline
          \textbf{Overall} & \bm{$81.2 \pm 1.3$} & 
          \bm{$84.0 \pm 1.4$ }\\ \hline
        \end{tabularx}
    \caption{There is little difference in average solution quality in percent with different initialisation values with 64 input vectors for different network sizes}
    \label{tab:init64}
\end{table}

\begin{table}[H]
    \centering
    \begin{tabularx}{0.8\linewidth}{|C|C|C|}
        \hline
        \rowcolor[gray]{0.9}
        \textbf{Locations} & \textbf{Input vector with zeros} & \textbf{Input vector with 0.5} \\ \hline
        4  & $100.0 \pm 0.0$ & $100.0 \pm 0.0$ \\ \hline
        5  & $100.0 \pm 0.0$ & $100.0 \pm 0.0$ \\ \hline
        6  & $100.0 \pm 0.0$ & $100.0 \pm 0.0$ \\ \hline
        7  & $100.0 \pm 0.0$ & $100.0 \pm 0.0$ \\ \hline
        8  & $100.0 \pm 0.0$ & $100.0 \pm 0.0$ \\ \hline
        9  & $99.0 \pm 0.9$  & $100.0 \pm 0.0$ \\ \hline
        10 & $96.6 \pm 0.6$  & $91.5 \pm 0.6$  \\ \hline
        11 & $91.9 \pm 0.9$  & $98.4 \pm 0.4$  \\ \hline
        12 & $84.4 \pm 2.3$  & $85.2 \pm 0.6$  \\ \hline
        15 & $71.3 \pm 0.4$  & $70.1 \pm 1.2$  \\ \hline
        17 & $77.3 \pm 3.1$  & $74.2 \pm 2.1$  \\ \hline
        26 & $89.7 \pm 0.9$  & $90.6  \pm 0.0$ \\ \hline
        42 & $100.0 \pm 0.0$ & $100.0 \pm 0.0$ \\ \hline
        48 & $10.2 \pm 0.2$ & $9.7 \pm 0.0$    \\ \hline
        \hline
        \textbf{Overall} & \bm{$87.2 \pm 0.7$} & 
        \bm{$87.1 \pm 0.4$} \\ \hline
    \end{tabularx}
    \caption{There is no significant difference in VQAi-CML between the average solution quality in percent for different initialisation values with 256 input vectors and different network sizes}
    \label{tab:init256}
\end{table}

\begin{landscape}

\begin{table}[h]
    \centering
    \begin{tabularx}{1.0\linewidth}{|p{1.5cm}|*{10}{C|}}
        \hline
        \rowcolor[gray]{0.8}
        \textbf{loc/input} &
        \textbf{2}   &
        \textbf{4}   &
        \textbf{8}   &
        \textbf{16}  &
        \textbf{32}  &
        \textbf{64}  &
        \textbf{128} &
        \textbf{256} &
        \textbf{512} &
        \textbf{1,024}
        \\
        \hline
         4 & 100.0 & 100.0 & 100.0 & 100.0 & 100.0 & 100.0 & 100.0 & 100.0 & 100.0 & 100.0\\
         \hline
         5 & 100.0 & 100.0 & 100.0 & 100.0 & 100.0 & 100.0 & 100.0 & 100.0  & 100.0 & 100.0\\
         \hline
         6 &  94.0 &  94.0 &  94.0 &  94.0 & 100.0 & 100.0 & 100.0 & 100.0  & 100.0 & 100.0\\
         \hline
         7 & 100.0 &  89.7 &  89.7 &  87.7 &  89.0 &  83.2 &  89.0 & 100.0  & 100.0 & 100.0\\
         \hline
         8 &  82.4 &  83.5 &  82.7 &  83.3 & 100.0 &  87.0 & 100.0 & 100.0  & 100.0 & 100.0\\
         \hline
         9 &  92.1 &  83.3 &  74.0 &  81.4 &  92.1 &  88.9 & 100.0 &  99.8  & 100.0 & 100.0\\
         \hline
        10 &  64.3 &  72.9 &  61.1 &  73.3 &  73.7 &  65.1 &  88.3 &  97.3  & 100.0 & 97.5\\
        \hline
        11 &  86.3 &  87.5 &  87.5 &  90.7 &  87.5 &  88.5 &  87.5 &  92.0  & 98.8 & 98.7\\
        \hline
        12 &  65.1 &  65.8 &  58.8 &  63.3 &  70.0 &  81.9 &  78.7 &  88.1  & 83.5 & 90.7\\
        \hline
        15 &  59.5 &  70.8 &  73.9 &  65.4 &  65.5 &  61.8 &  73.1 &  71.9  & 73.3 & 74.7\\
        \hline
        17 &  58.3 &  72.4 &  72.0 &  58.3 &  71.8 &  73.0 &  72.6 &  74.6  & 77.9 & 82.7\\
        \hline
        26 &  91.5 &  92.7 &  89.0 &  94.1 &  94.1 &  94.1 &  90.6 &  90.6  & 90.6& 90.6\\
        \hline
        42 & 100.0 & 100.0 & 100.0 & 100.0 & 100.0 & 100.0 & 100.0 & 100.0  & 100.0 & 100.0\\
        \hline
        48 &   9.2 &   8.6 &   9.0 &   8.8 &   8.9 &   9.5 &   9.5 &  10.1  & 9.5 & 10.3\\
        \hline
    \end{tabularx}
    \caption{The  solution quality in percent increases as the number of input vectors (mini-batches) increases for VQAi-CML.}
    \label{tab:bin_input}
\end{table}

\end{landscape}

\section{Appendix - Supplementary figures}
\label{appendix_figures}


\begin{figure}[h]
    \centering
    \includegraphics[width=1.0\linewidth]{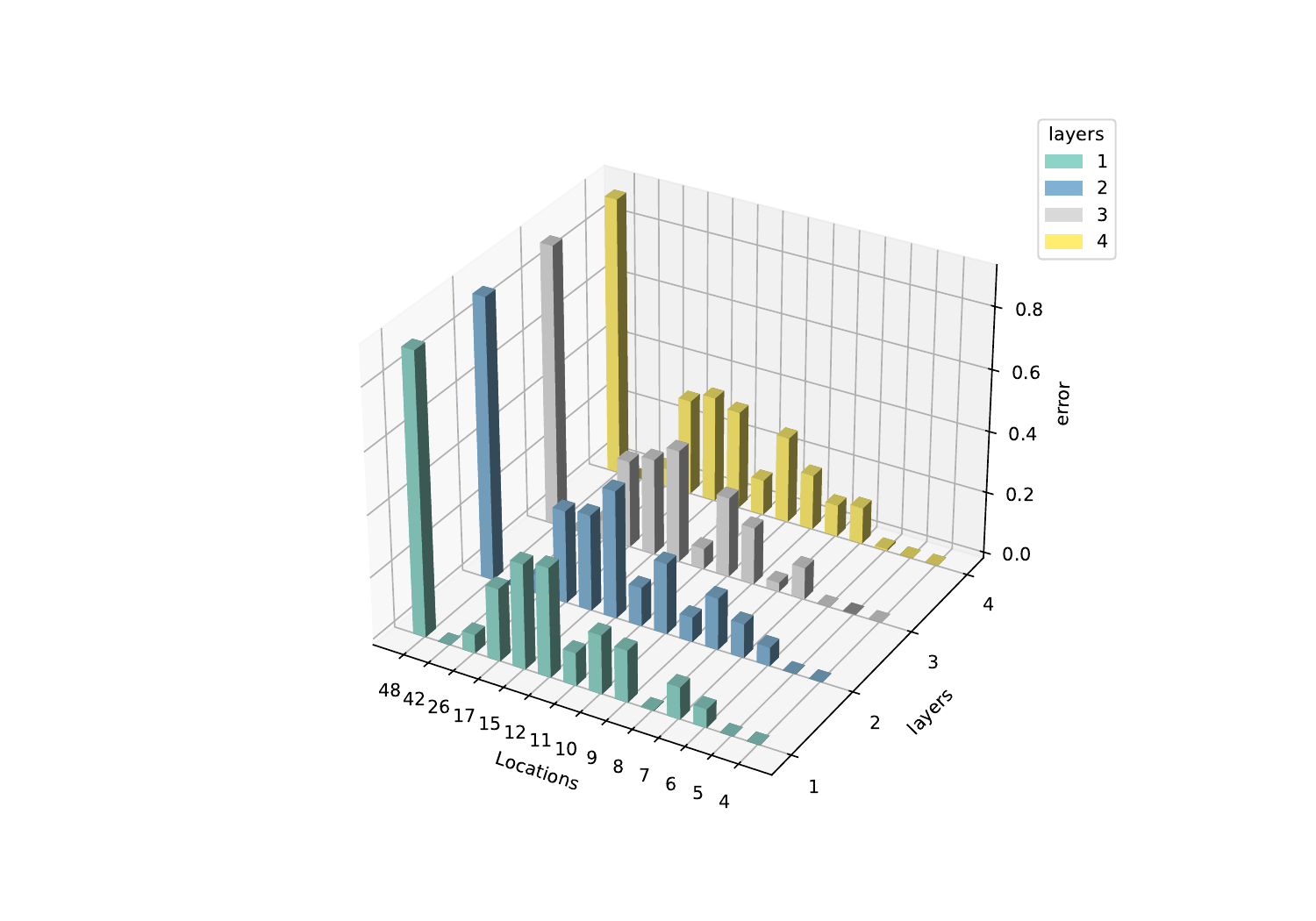}
    \caption{Error rates for VQAi-CML for different locations and layer, showing the number of layers makes little difference.}
    \label{fig:ML1}
\end{figure}

\end{document}